\documentclass[aps,pra,twocolumn,longbibliography,superscriptaddress,floatfix]{revtex4-1}

\usepackage{graphicx}% Include figure files
\usepackage{dcolumn}% Align table columns on decimal point
\usepackage{bm}% bold math
\usepackage{amsmath}
\usepackage{mathtools}
\usepackage{graphicx}
\usepackage{ulem}
\usepackage{epstopdf}
\usepackage{subfigure}
\usepackage{color}
\usepackage{amsthm}
\usepackage{newlfont}
\usepackage{graphicx}
\usepackage{epstopdf}
\usepackage{appendix}
\usepackage[breaklinks=true]{hyperref}
\usepackage{breakcites}
\usepackage{textcomp}
\usepackage{appendix}
\usepackage{multirow}	
\usepackage{color}
\usepackage{amssymb}
\usepackage{epsfig}
\usepackage{bm}
\usepackage[american]{babel}
\usepackage{braket}
\usepackage{soul}
\usepackage{float}
\hypersetup{
     colorlinks   = true,
     linkcolor    = red,
     citecolor    = cyan,
     urlcolor     = magenta
     }
\pdfoutput=1 

\begin{document}

% \preprint{APS/123-QED}

\title{Nonlinear Topological Edge States: from Dynamic Delocalization to Thermalization}% Force line breaks with \\
% \thanks{A footnote to the article title}%

% \author{Ann Author}
%  \altaffiliation[Also at ]{Physics Department, XYZ University.}%Lines break automatically or can be forced with \\
\author{Bertin Many Manda}%
%  \email{Second.Author@institution.edu}
\affiliation{LAUM, CNRS, Le Mans Universit\'e, Avenue Olivier Messiaen, 72085 Le Mans, France}
\affiliation{Nonlinear Dynamics and Chaos group, Department of Mathematics and Applied Mathematics, University of Cape Town, Rondebosch, 7701 Cape Town, South Africa}

\author{Rajesh Chaunsali}
\affiliation{LAUM, CNRS, Le Mans Universit\'e, Avenue Olivier Messiaen, 72085 Le Mans, France}
\affiliation{Department of Aerospace Engineering, Indian Institute of Science, Bangalore 560012, India}

\author{Georgios Theocharis}
\affiliation{LAUM, CNRS, Le Mans Universit\'e, Avenue Olivier Messiaen, 72085 Le Mans, France}

\author{Charalampos Skokos}
%  \homepage{http://www.Second.institution.edu/~Charlie.Author}
\affiliation{Nonlinear Dynamics and Chaos group, Department of Mathematics and Applied Mathematics, University of Cape Town, Rondebosch, 7701 Cape Town, South Africa}
% \affiliation{
%  Third institution, the second for Charlie Author
% }%
% \author{Delta Author}
% \affiliation{%
%  Authors' institution and/or address\\
%  This line break forced with \textbackslash\textbackslash
% }%

% \collaboration{CLEO Collaboration}%\noaffiliation

\date{\today}% It is always \today, today,
             %  but any date may be explicitly specified

\begin{abstract}
We consider a mechanical lattice inspired by the Su–Schrieffer–Heeger model along with cubic Klein-Gordon type nonlinearity. We investigate the long-time dynamics of the nonlinear edge states, which are obtained by nonlinear continuation of topological edge states of the linearized model. Linearly unstable edge states delocalize and lead to chaos and thermalization of the lattice. Linearly stable edge states also reach the same fate, but after a critical strength of perturbation is added to the initial edge state. We show that the thermalized lattice in all these cases shows an effective renormalization of the dispersion relation. Intriguingly, this renormalized dispersion relation displays a unique symmetry, i.e., its square is symmetric about a finite squared frequency, akin to the chiral symmetry of the linearized model. 

\end{abstract}

%\keywords{Suggested keywords}%Use showkeys class option if keyword
                              %display desired
\maketitle

%\tableofcontents

%%%%%%%%%%%%%%%%%%%%%%%%%%%%%%%%%%%%%%%%%%%%%%%%%%%%%%%%%
\section{\label{sec:intro}Introduction}

The problem of energy transport and ensuing \textit{thermalization in nonlinear discrete lattices} has attracted a lot of attention over the past decades, starting with the acclaimed numerical experiment of Fermi, Pasta, Ulam and Tsingou~\cite{FPUT1955,F1992,BI2005,CRZ2005} to its far reaching implications in understanding integrable and non-integrable dynamical systems with applications to statistical mechanics, ergodicity and chaos (see e.g.,~\cite{DLL1995,FIK2006,PCSF2011,BCP2013,CBTD2016,CGMGG2018}). 
One central and longstanding question is whether and how the intrinsic dynamics of a discrete nonlinear lattice leads to chaos, thermalization and energy equipartition across all the normal modes (NMs).
Several works have addressed this question for many different one-dimensional (1D) lattice models.
In~\cite{RCK2000}, dynamical regimes in the parameter space of the discrete nonlinear Schr\"odinger (DNLS) system leading to thermalization or to the appearance of localized motions were identified, while energy thresholds allowing the emergence of such localized structures were obtained in~\cite{R2008}. Furthermore, the chaotic dynamics of the model was investigated in more detail in~\cite{IP2021} by means of the Lyapunov exponents spectrum.
Studies of the disordered version of the DNLS system, where all linear modes are exponentially localized, showed that the nonlinear dynamics leads to energy delocalization~\cite{FKS2009}, chaos~\cite{SMS2018} and thermalization~\cite{MAPS2009}. The onset of thermalization and equipartition due to the nonlinear interactions between normal modes for the discrete nonlinear Klein-Gordon (KG) lattice was studied in~\cite{POC2018}, while the chaotic spreading of initially localized excitations in the disordered KG system was investigated in~\cite{SGF2013,SMS2018}.

In recent years, \textit{topological discrete lattices} have attracted a sheer amount of attention in theory and experiments~\cite{HK2010,OPA2019, MXC2019}.
The core concept involves designing systems with nontrivial band topology that results in boundary states robust to lattice's imperfection and/or smooth deformation.
While these so-called topological states are well defined in the linear limit, there has been an intense and increasing interest, mostly in the fields of photonics, electronics (see for example~\cite{SLC2020}) and mechanics~\cite{CUV2014,PVL2018,CT2019,VPR2019,DL2019,DY2019,LSCJRL2021,TMV2021}, to what happens when nonlinear effects enter into play. 
Depending on the functional form of the nonlinearity, the latter can modify the shape, the frequency and more importantly the stability of the topological edge modes~\cite{LRPS2016,CXYKT2021,LP2021,MS2021}. The presence of nonlinearity, can also lead to 
the formation of topologically-robust edge solitons~\cite{ACM2014}, unique gap solitons~\cite{LPRS2013}, or “self-induced” edge solitons and domain walls~\cite{HVA2017,CT2019,LC2021}. Finally, nonlinear, periodically driven topological lattices have been also investigated both theoretically and experimentally~\cite{ACM2014,BWPE2019,MKO2020,MR2020a,MR2020b,MMK2021}

%Especially in mechanical lattices, %where nonlinear functional form can %be designed by geometric %manipulations~\cite{JK2008, %FOJ2014}, it is instructive to explore several aspects of the interplay between nonlinearity and band topology~\cite{CUV2014, PVL2018, CT2019, VPR2019, DL2019, DY2019}. 
All these works contribute to our understanding of how topological states behave or emerge in the presence of inter-particle interactions and nonlinearity. 
However, very little is known about the long time behavior of these nonlinear topological states~\cite{LSMSS2021}. Even, when a linear stability analysis is performed~\cite{CXYKT2021,MS2021}, this is able only to describe the short time dynamics. Since, the nonlinearity in general breaks the integrability of the underlined discrete lattice dynamical system, equipartition or energy localization are the only two possibilities for long times.
%especially when one wants to connect 
%the dynamics of nonlinear topological lattices with the ones of thermalization and wave turbulence of nonlinear lattices. %

% The emergence of the two latter scenarios have been extensively investigated for several prototypical (generic) lattices models such as the Fermi-Pasta-Ulam-Tsingou (FPUT), discrete nonlinear Schr\"odinger (DNLS) and Klein-Gordon (KG) systems among others~\cite{F1992,BI2005,CRZ2005,G2007,RCK2000,PCBLO2019}.
% For example, considering we initially introduce a localized excitation on the lattice either in the real or normal mode (NM) spaces, localization in the long time limit will occur if the nonlinearity is very small (KAM regime) or sufficiently strong (selftrapping regime) such that the frequency of the nonlinear modes of the system are tuned out of resonance with.

Topological lattices possess robust edge states within a topologically protected band gap. 
Since this scenario is absent in generic nonlinear lattice models (see, e.g.~\cite{FPUT1955,F1992,BI2005}) a series of 
questions with a fundamental theoretical interest arise.
Do topological, nonlinear discrete lattices, reach thermalization? 
Can nonlinear topological edge states use as initial conditions on a topological lattice, remain localized under the presence of perturbations, or a dynamical delocalization is expected? 
Considering that the nonlinear topological lattice has reached thermalization, can one, along the lines of  
\cite{OVPL2015,LO2018}, describe the system by an \textit{effective} dispersion relation with some unique symmetry 
properties inherited from the dispersion relation of the linearized topological lattice?

Trying to answer these questions, we perform here a follow up systematic analysis of the long time dynamics of 
nonlinear edge states, that were analyzed in a mechanical lattice with KG-type nonlinearity~\cite{CXYKT2021}. 
In the linearized limit, this model is inspired from the well-celebrated Su–Schrieffer–Heeger (SSH) lattice~\cite{SSH1979}, in which topological protection of the edge states stems from chiral symmetry of the Hamiltonian~\cite{SH2016}. However, their difference lies in the fact that the band gap falls in finite (non-zero) frequency range in the former~\cite{CXYKT2021}, and hence, the edge states have vibratory nature.
It was shown in~\cite{CXYKT2021} that nonlinear continuations of topological edges states could pass through regions of linear stability and instability depending on the sign and the strength of the nonlinearity. 
%Some nonlinear edge states of this model spread towards the bulk in a short time due to instabilities, and there are also some states that remain localized under small perturbations due to linear stability~\cite{CXYKT2021}. 

Here, we investigate the \textit{long-time} chaotic dynamics of \textit{unstable} states to be able to answer the question if the system achieves equipartition in a long time. Moreover, we investigate if the linearly \textit{stable} edge states can also be delocalized due to strong perturbations, and eventually, lead to equipartition.
%In particular, motivated by the recent results on the short time dynamics of unstable topological nonlinear edge states~\cite{CXYKT2021} for which energy spreading toward the bulk was observed, we investigate in details, its long time, chaotic dynamics.
%\textbf{GT: We focuses directly only on the unstable topo modes, while the surprising and "strong" result of our ~\cite{CXYKT2021} paper, is the existence of linearly stable topo modes. So I believe, we have to put this into the game from the beginning. This can be maybe done by saying that even the linearly stable ones, for strong perturbations (or not) leads also to equipartition, blablabla ...}
%Consequently, it is natural to know whether such wave delocalization could lead to the thermalization of the entire lattice.
%
%Therefore, we begin by examining the thermalization of 1D topological lattices starting in the neighborhood of both stable %and unstable nonlinear topological edge states.
%We use a combination of observables in the model's phase, NMs and tangent spaces like the participation number of the %distribution of 
%\textbf{?? maybe the energy per site/particle?} on-site energy, the entropy of the modal energy and the maximal Lyapunov %exponent whose evolutions are generated via direct numerical simulations, in order to characterize the system for %different energy values and nonlinear strengths \textbf{by this we mean different signs? nonlinearity strength and energy %values are equivalent}.
%
We demonstrate that despite the presence of the band gap into the frequency spectrum, which may prevent part of the energy of the edge state to propagate away from the edges, \textit{linearly unstable}, 
topological nonlinear edge states seen in Ref.~\cite{CXYKT2021}, lead to thermalization of the lattice well described by a Gibbs distribution. Then, focusing on the \textit{linearly stable} topological nonlinear edge states seen in Ref.~\cite{CXYKT2021}, we discuss the condition under which these loose their stability and lead to thermalization too.
%as one expects the system's phase space to be a region of mixed regular and chaotic motion under a monotonic variation of the nonlinearity.
%NMs (sites) \textbf{?? why not just thermalization of the lattice?}, whose energy distribution in thermal equilibrium is 
In both cases, by calculating the renormalized dispersion relation for the thermalized state, we find, remarkably, that this possesses a unique symmetry akin to chiral symmetry seen in the dispersion of the linearized lattice.

%To conclude, we find that the idea of robust, nonlinear topological states, for the case of mechanical lattices with KG nonlinearity, is highly compromised in the limit of strong nonlinearity, as in this limit, the system settles out into a thermalized equilibrium state, where effectively it can again be described by a normalized dispersion relation with a chiral-like symmetry.

The paper is organized as follows. 
In Sec.~\ref{sec:model} we introduce the Hamiltonian formalism of the model in our study.
In Sec.~\ref{sec:topological_edge_states}, we present the method for obtaining linear edge states and their nonlinear continuations, discussing their spatial and spectral properties.  
In Sec.~\ref{sec:computational_method}, we present the computational foundations of our numerical investigation.
Section~\ref{sec:numberical_simulations} contains our numerical results, which are mainly focused on the energy spreading of unstable topological nonlinear edge states, renormalized dispersion of the thermalized state, as well as the robustness of stable edge states to increasing perturbations.
Finally in Sec.~\ref{sec:conclusion} we conclude our work and present some open avenues for future research.
%In appendix~\ref{app-sec:symplectic_integration} we provide details on the numerical integration scheme we implemented for our computer simulations.

%%%%%%%%%%%%%%%%%%%%%%%%%%%%%%%%%%%%%%%%%%%%%%%%%%%%%%%%%
\section{\label{sec:model}Model description}

\begin{figure}[h]
    \centering
    \includegraphics[width=1\columnwidth]{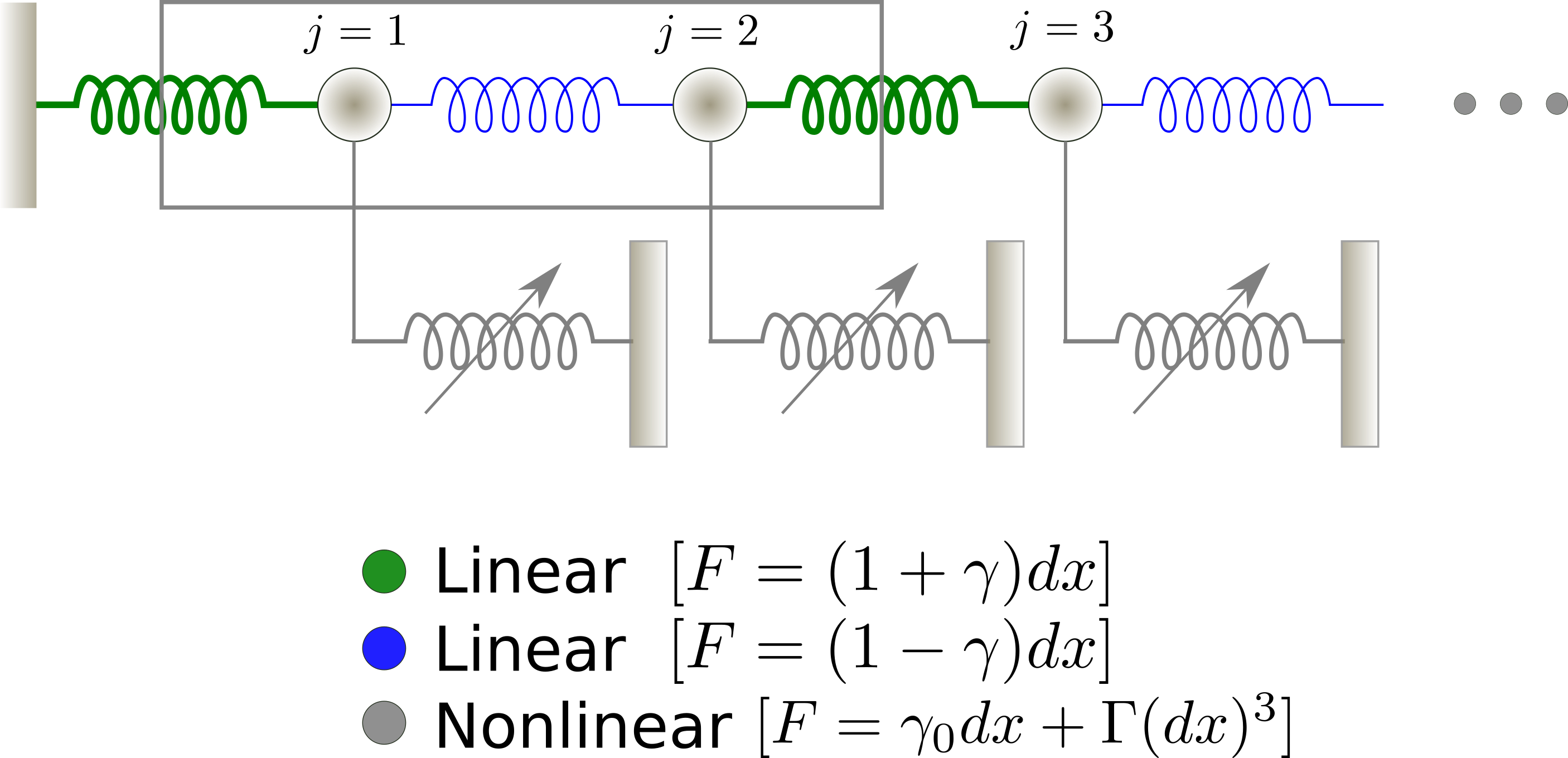}
    \caption{A semi-infinite chain consisting of equal masses [grey spheres] interconnected with two types of linear springs with elastic constants $1+\gamma$ and $1-\gamma$, denoted by the green and blue spiral coils respectively. Each mass is grounded with a nonlinear spring [grey spiral coil] whose elastic and nonlinear constants are $\gamma_0$ and $\Gamma$. 
    The $dx$ denotes the spring deformation and $F$ the resulting external force applying to a single mass.
    %We look for the long-time dynamics of the state localized on the left edge of the chain.
    }
    \label{fig:description_ssh_kg}
\end{figure}

Our system is a lattice of $n$ coupled identical particles (see Fig.~\ref{fig:description_ssh_kg}). These particles are interconnected with alternating binary springs of strengths $1-\gamma$ and $1+\gamma$, and are subject to onsite nonlinear potentials whose elastic and nonlinear coefficient values are given by $\gamma_0$ and $\Gamma$ respectively~\cite{CXYKT2021}. In the limit $\gamma \rightarrow 0$ we recover the %`generic' 
monomer KG chain~\cite{S2011,POC2018}.
As an additional remark, we note that when $\Gamma < 0$ the system possesses escape energy thresholds.
The total energy of a lattice is given by the Hamiltonian function
\begin{equation}
    \begin{split}
    &\mathcal{H} = \mathcal{H}_2 + \mathcal{H}_4, \quad \mathcal{H}_4 = \frac{\Gamma}{4} \sum_{j=1}^{n} x_j^4, \\
    &\mathcal{H}_2=\frac{1}{2} \sum_{j=1}^{n} \left[p_j^2 + \gamma_0 x_j^2 \right] \\
        & + \frac{1}{4} \sum_{\substack{j=1 \\ j = \text{odd}}}^{n} \left[(1 + \gamma) \left(x_j - x_{j - 1}\right)^2 + (1 - \gamma)\left(x_{j + 1} - x_j\right)^2 \right] \\
        & + \frac{1}{4} \sum_{\substack{j=0 \\ j = \text{even}}}^{n} \left[ (1 - \gamma) \left(x_j - x_{j - 1}\right)^2 + (1 + \gamma)\left(x_{j + 1} - x_j\right)^2 \right],
    \end{split}
    \label{eq:hamilton_tkg}
\end{equation}
where, $x_j$ and $p_j$ are the displacements from 
the equilibrium points, and the associated momentum of the $j$th particle. It should be emphasized that the quartic term $\mathcal{H}_4$ is of KG type nonlinearity. Further, as we are going to see in Sec.~\ref{sec:topological_edge_states}, under proper boundary conditions, the Hamiltonian model [Eq.~\eqref{eq:hamilton_tkg}] allows for (non)linear topological edge states.

The lattice's equations of motion are derived from the Hamiltonian function $\mathcal{H}$ [Eq.~\eqref{eq:hamilton_tkg}] taking into account that the pair $x_j, ~p_j$ are canonical conjugate variables
\begin{equation}
    %\left.
        \begin{split}
            \dot{x}_j &= p_j, \\
            \dot{p}_j &= - \left(1 + \gamma\right)\left(x_j - x_{j - 1}\right) + \left(1 - \gamma\right)\left(x_{j + 1} - x_j\right)  \\
            & \qquad - \gamma_0 x_j - \Gamma x_j^3 \quad \mbox{if $j$ odd}, \\
            \dot{p}_j &= - \left(1 - \gamma\right)\left(x_j - x_{j - 1}\right) + \left(1 + \gamma\right)\left(x_{j + 1} - x_j\right) \\
            & \qquad - \gamma _0 x_j - \Gamma x_j^3 \quad \mbox{if $j$ even},
        \end{split} 
    %\right\}
    \label{eq:eqs_of_motion_tkg}
\end{equation}
with $(\,\dot{ }\,)$ denoting the time derivative.
The full set of equations [Eq.~\eqref{eq:eqs_of_motion_tkg}] conserves the value of the Hamiltonian function [Eq.~\eqref{eq:hamilton_tkg}] which could therefore be used as our control parameter of the nonlinearity strength when $\gamma_0$, $\gamma$ and $\Gamma$ remain fixed.

A common way to investigate the system's thermalization is through the dynamics of its NMs~\cite{MAPS2009,S2011,DCF2017,PCBLO2019}.
The NM coordinate is defined by~\cite{GLC2005,OVPL2015,LO2018}
\begin{equation}
    a_k = \frac{1}{\sqrt{2\omega_k \left(\gamma\right)}} \left[P_k + i\omega_k\left(\gamma\right) Q_k\right],
    \label{eq:normal_mode_variables}
\end{equation}
where $Q_k$, $P_k$ are the discrete Fourier transform of the canonical coordinates $x_j$ and momenta $p_j$, and $\omega_k \left(\gamma\right)$ 
%[Eq.~\eqref{eq:disp_relation}] 
is the frequency of mode $k$. For the case of periodic boundary conditions, the two bands of the dispersion relation are given by: 
%%which corresponds to $\omega_k\left(\gamma=0\right) = \sqrt{\gamma_0+4\sin^2\left(k\pi/\textcolor{red}{2} n\right)}$ when $\gamma =0$~\cite{POC2018,PCBLO2019} (monomer KG model).
\begin{equation}
    \omega_k= \left( 2 + \gamma_0 \pm \sqrt{2(1+\gamma^2) + 2(1-\gamma^2) \cos{(2 k\pi/n)}} \right)^{1/2}.
    \label{eq:disp_relation}
\end{equation}
Thus, $\gamma$ adjusts the size of the band gap in the dispersion relation (see Sec.~\ref{sec:topological_edge_states}).
In the normal variables, the Hamiltonian function $\mathcal{H}$ [Eq.~\eqref{eq:hamilton_tkg}] is written as 
\begin{equation}
    \begin{split}
        \mathcal{H} = &\hat{\mathcal{H}}_2 + \hat{\mathcal{H}}_4, \quad \hat{\mathcal{H}}_2 = \sum _{k=0}^{n} \omega_k \left(\gamma\right) \lvert a_k \rvert^2, \\
        &\hat{\mathcal{H}}_4 = \Gamma V(a_1, a_2, \ldots, a_n),
    \end{split}
    \label{eq:hamilton_tkg_fourier}
\end{equation}
where $V(a_1, a_2, \ldots, a_n)$ is a linear combination of quartic products of  $a_k$ and $a_k^{\star}$~\cite{POC2018,PCBLO2019}.
Consequently the equations of motion become $i\dot{a}_k = \partial \mathcal{H}/\partial a_k^\star$.
Here the $( {}^{\star} )$  denotes the complex conjugate.
%It is worth emphasizing that throughout its evolution, system [Eq.~\eqref{eq:eqs_of_motion_tkg}] does not conserve $\hat{\mathcal{H}}_2$ when $\Gamma \neq 0$.

% \textbf{GT: Do we use equation 4 or not? In case that this equation and the relavant discussion below are not used, for simplicity and clarity we could take it out... to be checked}
% {\color{red}BMM: There are some mention of $\hat{\mathcal{H}}_2$ here and there along the Secs. 4 and 5.
% But I guess we can survive without it by just changing all the $\hat{\mathcal{H}}_2\rightarrow \mathcal{H}_2$ into the text.
% }
%%%%%%%%%%%%%%%%%%%%%%%%%%%%%%%%%%%%%%%%%%%%%%%%%%%%%%%%%
\section{\label{sec:topological_edge_states}Topological (non-)linear edge states}
\begin{figure}[h]
    \centering
    \includegraphics[width=\columnwidth]{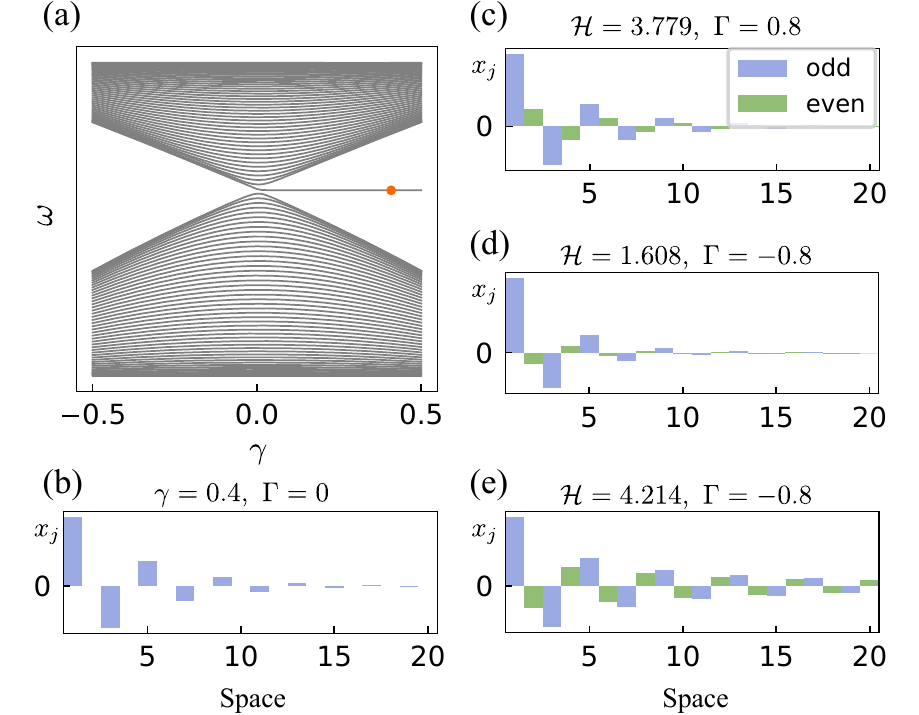}
    \caption{Nonlinearity-modified topological edge states. (a) Emergence of topological edge state inside the band gap. (b) Spatial profile, $x_j$ of the edge state with $\gamma_0=1$ and  $\gamma=0.4$ whose frequency is indicated by the red dot in panel (a). (c)--(e) Nonlinear continuations of the topological edge state for 
    $(\mathcal{H}=3.779, \Gamma=0.8)$, 
    $(\mathcal{H}=1.608, \Gamma=-0.8)$ and $(\mathcal{H}=4.214, \Gamma=-0.8)$, respectively. 
    In all the cases, the chosen phase of the mode is such that the corresponding momentum profile is zero, $p_j$=0.
    }
    % \textbf{GT: In all the cases, the chosen phase of the mode is such that the corresponding momentum profile is zero, $p_j$=0}.}
    \label{fig:linear_disper_rel_&_topological_states_spatial_profile}
\end{figure}

Since our system is the one studied in Ref.~\cite{CXYKT2021}, in this section we provide a very brief review of essential details on the characteristics of topological states to be useful in the current study. 
We can get the first insight on the nature of our system by evaluating its dispersion properties in the linear limit ($\Gamma=0$) through the solution of an eigenvalue problem of a dynamical matrix. This matrix is akin to the Hamiltonian of the SSH model~\cite{SSH1979} apart from the fact that its diagonal elements are nonzero. 
Consequently, the eigenvalues ($\omega^2$), are symmetric about the mid gap frequency $\omega^2 = 2 + \gamma_0$ due to chiral symmetry of the dynamical matrix~\cite{SH2016}.  
The symmetry makes it possible to characterize the dynamical matrix with a topological invariant, e.g., winding number. The system can be shown to make a topological transition when the parameter $\gamma$ varies from negative to positive values, or in other words, band gap closes and opens again~\cite{SH2016}.

According to the bulk-boundary correspondence, the topological transition of the \textit{infinite} lattice as discussed above reflects as the emergence of edge states in the \textit{finite} chain~\cite{HK2010}. 
%Since we have considered a semi-infinite chain, we expect a topological state localized on the left edge of the chain shown in Fig.~\ref{fig:description_ssh_kg}. 
For all the cases in this work, we take a long chain of $n=100$ particles, to avoid strong finite size effects. We also consider fixed (free) boundary condition for the left (right) end of the chain. This choice ensures the existence of only one edge state localized around the fixed boundary. We note that in the case of even number of particles and with fixed boundary conditions in both ends, two hybridized edge states are expected, see for example \cite{SKCATY2021}. In Fig.~\ref{fig:linear_disper_rel_&_topological_states_spatial_profile}(a), we show the spectrum of this finite chain when $\gamma_0 = 1$. Due to the topological transition at $\gamma=0$, we observe the emergence of a state pinned at $\omega^2 = 2 + \gamma_0$ inside the band gap for $\gamma>0$. This state is exponentially localized on the left edge, as shown in Fig.~\ref{fig:linear_disper_rel_&_topological_states_spatial_profile}(b) for $\gamma=0.4$. Moreover, the state has zero amplitude at even sites as a result of chiral symmetry. Such a state is of topological origin and protected by chiral symmetry. 

Once nonlinearity is turned on, i.e., $\Gamma \neq 0$ in $\mathcal{H}$ [Eq.~\eqref{eq:hamilton_tkg}], we calculate the time-periodic solutions using Newton's method in the system's phase space (see e.g.~\cite{FG2008}) with the aforementioned edge state as the initial condition. Moreover, we determine the linear stability of such nonlinear states using Floquet theory~\cite{A2006}.
In Ref.~\cite{CXYKT2021} it was shown that the topological edge state changes its frequency, shape, and stability depending on the sign of nonlinearity and increasing its energy, thus the strength of nonlinearity. We show a few representative cases in Figs.~\ref{fig:linear_disper_rel_&_topological_states_spatial_profile}(c)--(e). These are not only distinct from the linear edge state shown in Fig.~\ref{fig:linear_disper_rel_&_topological_states_spatial_profile}(b) in that they no longer retain the chiral profile with vanishing amplitudes at even sites, but also they have  different instabilities. For example, the stiffening ($\Gamma>0$) case shown in Fig.~\ref{fig:linear_disper_rel_&_topological_states_spatial_profile}(c) with 
$(\mathcal{H}=3.779, \Gamma=0.8)$ is a linearly stable solution.
%``bulk-bulk'' instability due to the collision of two \textit{bulk} modes (both spatially extended) in the phase of Floquet multipliers~\cite{CXYKT2021}. 
Whereas the softening ($\Gamma<0$) cases shown  in Figs.~\ref{fig:linear_disper_rel_&_topological_states_spatial_profile}(d) and \ref{fig:linear_disper_rel_&_topological_states_spatial_profile}(e) with 
$(\mathcal{H}=1.608, \Gamma=-0.8)$ and $(\mathcal{H}=4.214, \Gamma=-0.8)$, respectively, are unstable solutions. 

%%%%%%%%%%%%%%%%%%%%%%%%%%%%%%%%%%%%%%%%%%%%%%%%%%%%%%%%%
 \section{\label{sec:computational_method}Computational framework}
\subsection{\label{subsec:equipartition_threshold} Equipartition}
In order to clarify the thermal properties of our topological lattice, we follow the observables of the system in the NM and real phase spaces.
% \textbf{GT: you refer to Lyaponov here? Is not clear what a tangent space is and in this subsection, we talk only about Fourier and real space}.
In the Fourier (NM) space, we follow the time evolution of the normalized modal energy~\cite{DCF2017}
\begin{equation}
    \nu_k (t) = \frac{E_k(t)}{\hat{\mathcal{H}}_2(t)}, \quad E_k(t) = \omega _k \left(\gamma\right) \lvert a_k (t) \rvert^2,
    \label{eq:modal_energy}
\end{equation}
where $\hat{\mathcal{H}}_2(t) =\sum_k E_k(t)$ is the total quadratic energy of the modes of Eq.~\eqref{eq:hamilton_tkg_fourier} and $E_k$ the energy of the $k$th mode.
We characterize the degree of spreading of $\nu_k$ in Eq.~\eqref{eq:modal_energy} using the spectral entropy~\cite{LPRSV1985,PL1990,P1991,GLL1992} 
\begin{equation}
    S(t)=-\sum_k \nu_k (t) \ln \left[ \nu_k(t)\right], \quad 0 \leq S(t) \leq \ln n.
    \label{eq:spectral_entr}
\end{equation}
If the total energy $\hat{\mathcal{H}}_2$ [Eq.~\eqref{eq:hamilton_tkg_fourier}] is localized on a single mode, $S=0$. 
On the other hand, if the energy $\hat{\mathcal{H}}_2$ is uniformly distributed across all modes, $S =\ln n$.
In order to mitigate the dependence on the lattice size $n$ of $S$ [Eq.~\eqref{eq:spectral_entr}], it is more convenient to use the rescaled spectral entropy~\cite{LPRSV1985,PL1990,P1991,GLL1992}
\begin{equation}
    \eta (t) = \frac{S(t) - S_{max}}{S(0) - S_{max}}, \quad 0 \leq \eta (t) \leq 1,
    \label{eq:resc_entr}
\end{equation}
with $S_{max}=\ln n$.
In this way, 
%the highly inhomogeneous lattice 
when the whole energy is concentrated at one mode $\eta=1$, while $\eta = 0$ corresponds to a uniform distribution of the energy throughout all the modes.
%Interestingly enough, 
If we consider that the modal energies at thermal equilibrium are characterized by a Gibbs distribution 
\begin{equation}
  \rho_{G,k} = \frac{1}{\mathcal{Z}}e^{-\beta E_k}, \quad \mathcal{Z} = \sum _k e^{-\beta E_k},
  \label{eq:gibbs_functional}
\end{equation}
where $\rho_{G, k}$ corresponds to the probability to find a mode $k$ with energy $E_k$, $\mathcal{Z}$ equals the partition function and $\beta$ is the inverse temperature,
% \textbf{GT: what is $\rho_{G,k}$, what is Z} 
we find that the value of $\eta(t)$ [Eq.~\eqref{eq:resc_entr}] at thermal equilibrium fluctuates about the average~\cite{GLL1992} 
\begin{equation}
  \langle \eta \rangle_G = \frac{1 - C}{S_{max} - S(0)}, 
  \label{eq:thermodynamic_resc_entr}
\end{equation}
where $C \approx 0.5772$ is the Euler constant.

To account for the phase space dynamics, we compute the normalized energy per site
\begin{equation}
    \xi_j (t) = \frac{h_j(t)}{\mathcal{H}}, \quad \mathcal{H} = \sum_j h_j(t),
    \label{eq:normalized_ener_per_site}
\end{equation}
where the energy at the $j$th site in the bulk is given by
\begin{equation}
    \begin{split}
     h_{j}  &= p_j^2/2 + \gamma_0 x_j^2/2 + \Gamma x_j^4/4 + (1 + \gamma) (x_j - x_{j - 1})^2/4 \\ 
    & + (1 - \gamma)(x_{j + 1} - x_j)^2/4 \quad \mbox{if $j$ odd}, \\
    h_{j} &= p_j^2/2 + \gamma_0 x_j^2/2 + \Gamma x_j^4/4 + (1 - \gamma) (x_j - x_{j - 1})^2/4 \\
    & + (1 + \gamma)(x_{j + 1} - x_j)^2/4 \quad \mbox{if $j$ even}.
    \label{eq:energy_at_site}
    \end{split}
\end{equation}
For the edge sites respectively with indices $j=1$  and $j=n$, we have
\begin{equation}
    \begin{split}
        h_{1}  &= p_1^2/2 + \gamma_0 x_1^2/2 + \Gamma x_1^4/4 + (1 + \gamma) x_1^2/4 \\ 
    & + (1 - \gamma)(x_{2} - x_1)^2/4, \quad \mbox{and}\\ 
    h_{n}  &= p_n^2/2 + \gamma_0 x_n^2/2 + \Gamma x_n^4/4 + (1 + \gamma) (x_n - x_{n - 1})^2/4,
    \end{split}
\end{equation}
% at the first ($i=1$) and last ($i=n$) sites respectively
considering fixed boundary condition for the left ($x_0=0$) and free boundary condition for the right ($x_{n+1}=x_n$) end of the chain.
%when the boundary conditions of the chain in Fig.~\ref{fig:description_ssh_kg} are set in the Hamiltonian function %[Eq.~\eqref{eq:hamilton_tkg}].
Then, the degree of inhomogeneity of the spatial  distribution of the energy is well characterized by the participation number 
\begin{equation}
    P^{-1}(t) = \sum _{j} \xi_j^2(t), \quad 1 \leq P(t) \leq n.
    \label{eq:participation_number_ener_distr}
\end{equation}
When the total energy  $\mathcal{H}$ is concentrated only on one excited site, $P=1$ while in case of equipartition of the energy among all sites, $P=n$.

%%%%%%%%%% FIG %%%%%%%%%
\begin{figure*}[t]
    \centering
    \includegraphics[width=1\textwidth, height=0.275\linewidth]{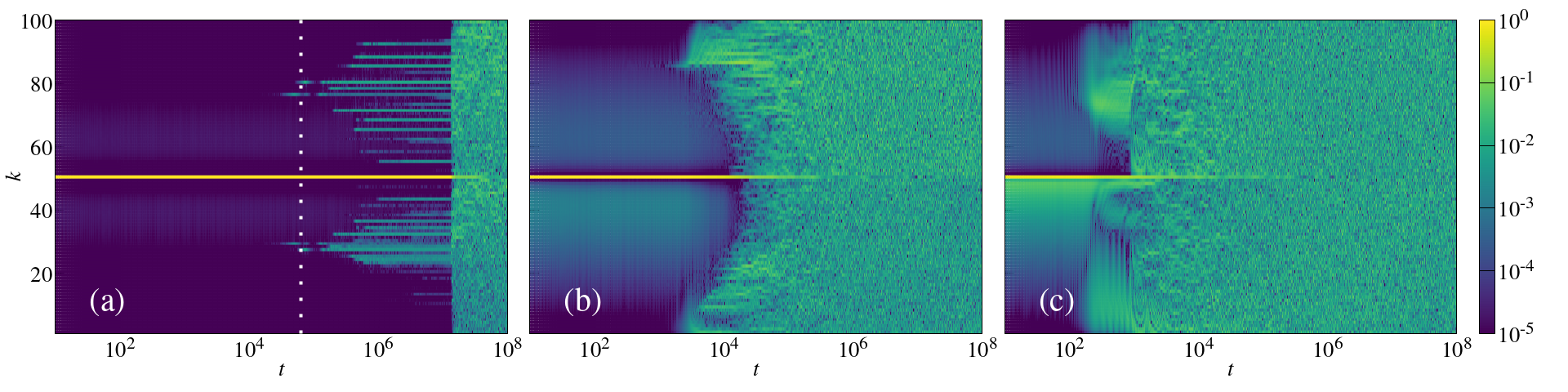}
    \caption{The normalized modal energy $\nu_k(t)$ [Eq.~\eqref{eq:modal_energy}] evolution profiles for representative realizations of setup [Eq.~\eqref{eq:init_state_localized}] of three different nonlinear topological edge states at ($\mathcal{H}$, $\Gamma$) values with ($\mathcal{H} = 0.307$, $\Gamma=-0.8$), ($\mathcal{H} = 1.608$, $\Gamma=-0.8$) and ($\mathcal{H}=4.214$, $\Gamma=-0.8$) in panels (a), (b) and (c) respectively. 
    The white vertical line in (a) shows the position of the cross section of $\nu_k (t)$ at $t\approx6.3\times 10^4$ depicted in Fig.~\ref{fig:floquet_multipliers} of App.~\ref{app-sec:floquet_resonance}.
    The edge states associated with panels (b) and (c) are respectively shown in Figs.~\ref{fig:linear_disper_rel_&_topological_states_spatial_profile}(d) and (e).
    Each point is colored according to the color scale at the right end of the figure.
    %   {\color{red}Change fig 3(a) with a $\Gamma=-0.8$ and append the text accordingly.}
    }
    \label{fig:distr_colormap_energy_mode_unstable}
 \end{figure*}
%%%%%%%%%%%%%%%%%%%%%%%%

%%%%%%%%%%%%%%%%%%%%%%%%%%%%%%%%%%%%%%%%%%%%%%%%%
\subsection{\label{subsec:stochastic_threshold}Chaotization}
As an additional characterization of the lattice's equipartition state, we compute the finite-time maximum Lyapunov exponent (ftMLE)~\cite{BGS1976,BGGS1980a,BGGS1980b,S2010}
\begin{equation}
    \lambda(t) = \frac{1}{t}\ln \left( \frac{\lVert \boldsymbol{W}(t) \rVert}{\lVert \boldsymbol{W}(0) \rVert} \right),
    \label{eq:ftmle}
\end{equation}
which quantifies the exponential growth rate at time $t$ of the separation $\boldsymbol{W}(t) $ in the phase space between two initially nearby orbits, so that $\lVert \boldsymbol{W}(0) \rVert \rightarrow 0$, with $\lVert \cdot \rVert$ being the usual Euclidean norm.
More specifically, the $\boldsymbol{W}(0) \allowbreak = \delta \boldsymbol{X}(0) \allowbreak = \left(\delta x_1(0), \delta x_2(0), \allowbreak \ldots, \delta x_n(0), \allowbreak \delta p_1(0), \allowbreak \ldots, \delta p_n(0) \right)$ can be viewed as a small perturbation to the system's initial position $\boldsymbol{X} (0)= \allowbreak \left(x_1(0), x_2(0), \ldots \allowbreak, x_n(0), p_1(0), \allowbreak \ldots, p_n(0)\right)$ in the phase space, 
% \textbf{GT: maybe initial vector? since it includes momenta too?} 
whose time evolution is governed by equations (referred to as variational equations)  derived from the linearization of the system's Hamilton equations of motion~\cite{SG2010}.
Consequently the maximal Lyapunov exponent (MLE)~\cite{BGGS1980a,BGGS1980b,S2010} 
\begin{equation}
    \Lambda = \lim_{t \rightarrow \infty} \lim _{\lVert \boldsymbol{W}(0) \rVert \rightarrow 0} \lambda(t),
    \label{eq:mle}
\end{equation}
measures the strength of chaos in the system.
% In addition, its inverse $T_\Lambda = 1/\Lambda$ (the so-called Lyapunov time) provide a characteristic time scale for the chaotization of the system, by practically quantifying the  time after which it becomes chaotic. \textcolor{green}{[HS: This paragraph needs improvement.]}

In thermal equilibrium, all the modes are heated and randomly interact in order to spatially homogenize chaos inside the lattice~\cite{DRT1997,SGF2013,SMS2018,MSS2020}. Therefore equipartition is associated to a constant positive ftMLE~\cite{BGS1976,P1991,F1992,CCPC1997,DRT1997}.
% In appendix B we underline the relation between the MLE and the Floquet multipliers of the spectral analysis performed in~\cite{CXYKT2021}.

% Characteristic of all these observables at equilibrium. Note that here equipartition means that the system is fully chaotic, stochastic or randomized.

 %%%%%%%%%%%%%%%%%%%%%%%%%%%%%%%%%%%%%%%%%%%%%%%%%%%%%%%%%
\section{\label{sec:numberical_simulations}Numerical simulations}
%%%%%%%%%%%%%%%%%%%%%%%%%%%%%%%%%%%%%%%%%%%%%%%%%%%%%%%%%%
In this section we present the numerical results obtained by evolving several perturbations of nonlinear topological localized states when both the energy and the sign of the onsite nonlinear strength are varied.
The numerical integration of the equations of motion and the variational equations of the model [Eq.~\eqref{eq:hamilton_tkg}] is done by the implementation of the $\mathcal{ABA}864$  symplectic scheme of order $4$ (see App.~\ref{app-sec:symplectic_integration} for details). 
The final integration time is $T \approx 10^{6}-10^{9}$, and  we set the integration time step to $\tau = 0.2$ so that the relative energy error, $E_r (t) = \lvert [\mathcal{H} (t) - \mathcal{H}(0)] / \mathcal{H}(0) \rvert$, remains below $E_r = 10^{-5}$ for the duration of all our simulations.

For all the cases presented in this work, the lattice size is fixed at $n=100$ particles, and the elastic parameters kept to $\gamma_0=1$ and $\gamma=0.4$. The choice of these elastic parameters leads to a relatively large band gap, Fig.~\ref{fig:linear_disper_rel_&_topological_states_spatial_profile}(a). 
%In addition, the generated harmonics of the nonlinear edge states, are located in the upper gap of the system, above the optical branch.
Investigating the cases of different values of $\gamma$, which controls the width of the frequency gap, is an interesting problem which however, is out of the scope of this work.
As we noted above, fixed and free boundary conditions are used for the left and right end of the chain, a choice which ensures the existence of only one edge state located around the left end of the chain.

Let us discuss now in some detail the initial conditions we use in our simulations. In the Fourier space representation, the topological edge state $\boldsymbol{X}^{b}=(x_1^{b}, x_2^{b}, \ldots, x_n^{b}, p_1^{b}, p_2^{b}, \ldots, p_n^{b})$ has coordinates $(Q_1^{b}, Q_2^{b}, \ldots, Q_n^{b}, P_1^{b}, P_2^{b}, \ldots, P_n^{b})$ such that in NM variables, $\boldsymbol{X}^{b}=(a_1^{b}, a_2^{b}, \ldots, a_n^{b})$ with $a_k^{b} = (P_k^{b} + i\omega _k (\gamma) Q_k^{b})/\sqrt{2\omega_k(\gamma)}$ [Eq.~\eqref{eq:normal_mode_variables}]. 
% \textcolor{green}{[HS: Why use $^{b}$? Do we also have $^{(a)}$. Probably we have to simplify our notation to only $^b$.]}
$\omega_k(\gamma)$ corresponds to the eigenfrequecies of the finite chain with the above mentioned boundary conditions and elastic parameters. We calculate these numerically by diagonalizing the system's dynamical matrix.
We initially start  with the topological localized state on which a real random phase $\varepsilon_k$ is added in its Fourier representation 
% \textcolor{green}{[HS: is also $\varepsilon \ll 1$? i.e. very small?]}{\color{red}BMM: We are already using the word `small' in the sentence.}
\begin{equation}
    a_k (t=0) = a_k^{b} e^{i\varepsilon_k}, \quad \mbox{if $1 \leq k \leq n$},
        \label{eq:init_state_localized}
\end{equation}
% {\color{red}BMM: move from $\varepsilon \rightarrow \varepsilon_k$ }
such that all the derived initial conditions possess the same linear energy as the edge state $\boldsymbol{X}^{b}$.
Unless otherwise stated, we use a uniform probability distribution to set the perturbation parameters $\varepsilon_k$ with values on the interval $[-10^{-2}, 10^{-2}]$. 

In order to compute the  ftMLE we choose  the coordinates of the initial  deviation vector $\boldsymbol{W}(t=0)$ as randomly selected numbers drawn from a uniform distribution in the interval $[-1,1]$. The non-zero coordinates of $\boldsymbol{W}(0)$  are located only inside the localization volume~\cite{LV0000} of the linear topological edge state~\cite{CXYKT2021} (see also Sec.~\ref{sec:topological_edge_states}).
It is worth mentioning that the final value of the long-time evolution of the ftMLE [Eq.~\eqref{eq:ftmle}] used to estimate the MLE [Eq.~\eqref{eq:mle}] does not practically depend on the choice of the initial deviation vector~\cite{S2010}.
% \textcolor{green}{[HS: We have to be careful and consistent in our statements: we do not compute the MLE, we compute the ftMLE and in this way we estimate the MLE.]}

 %%%%%%%%%%%%%%%%%%%%%%%%%%%%%%%%%%%%%%%%%%%%%%%%%%%%%%%%%%
 \subsection{\label{subsec:charact_unstable_topo_states}Characteristics of unstable topological edge state delocalization}
 
To provide a feeling of the thermalization of the lattice [Eq.~\eqref{eq:hamilton_tkg}] starting from unstable nonlinear topological edge states, we consider the sets of parameters $(\mathcal{H}=0.307, \Gamma=-0.8)$, $(\mathcal{H}=1.608, \Gamma=-0.8)$ and $(\mathcal{H}=4.214, \Gamma=-0.8)$ of the setup [Eq.~\eqref{eq:init_state_localized}].
The two latter correspond to the edge states shown in Figs.~\ref{fig:linear_disper_rel_&_topological_states_spatial_profile}(d)--\ref{fig:linear_disper_rel_&_topological_states_spatial_profile}(e).
We set the perturbation parameters $\varepsilon_k=\pm 10^{-2}$.
The amplitude profiles of the distribution of the normalized energy per mode $\nu_k(t)$ [Eq.~\eqref{eq:modal_energy}] is shown for representative realizations in Fig.~\ref{fig:distr_colormap_energy_mode_unstable}.
% Interestingly enough 
% \textbf{GT: not sure if it is so unexcepted to call interestingly enough} 
For all values of $\mathcal{H}$ [Eq.~\eqref{eq:hamilton_tkg}], the delocalization of unstable nonlinear topological edge states leads to the thermalization of the lattice at finite time scales. % although through different processes.

For example, in Fig.~\ref{fig:distr_colormap_energy_mode_unstable}(a), the normalized energy profile of the case $(\mathcal{H}=0.307, \Gamma=-0.8)$ is shown. This initial condition corresponds to a weakly unstable nonlinear edge state with ``bulk-bulk'' instability, which is caused by the collision of two \textit{bulk} modes (both spatially extended)~\cite{CXYKT2021}. It is known~\cite{MA1998} that these instabilities are caused by finite-size effects and that they are vanishing in the thermodynamic limit.
After a transient period of the order of $\approx 10^{4}$ time units, only $4$ modes in the acoustic ($2$ modes) and optical ($2$ modes) bands initially resonate.
We have checked that these initially excited modes of Fig.~\ref{fig:distr_colormap_energy_mode_unstable}(a), correspond to the frequencies of the most unstable eigenvectors of the Floquet analysis~\cite{CXYKT2021}, see App.~\ref{app-sec:floquet_resonance}.
Afterward, a cascading effect of resonances between NMs is taking place up to complete thermalization of the lattice at about $10^{7}$ time units.
%Interestingly enough, 
%
%of the presence of weakly chaotic dynamics in the system characterized by the slow relaxation of the system's initial condition, with the weakly unstable Floquet multipliers, namely the one with a small divergence from the unit circle~\cite{DRT1997,AB2006,CXYKT2021}. 
We have also checked that by increasing the size of the lattice to $1000$ particles, the instability and the resulting thermalization appears later, confirming the finite-size origin of this weak instability.
%confirming the fact that this weak, finite-size instability disappear in the thermodynamic limit.
%It is worthwhile stressing that the process depicted in Fig.~\ref{fig:distr_colormap_energy_mode_unstable}(a) may correspond to the near discrete exact resonances responsible of lattice's thermalization at small energy  $\mathcal{H}$ as discussed in the wave turbulence framework of discrete nonlinear lattice models (see e.g, Refs.~\cite{OVPL2015,LO2018,PCBLO2019}).

% \textbf{GT: This case corresponds to a weakly unstable nonlninear edge mode, with bulk-bulk instabilities, see ~\cite{CXYKT2021}. Interestingly enough, we have checked that some of the excited modes of Fig.~\ref{fig:distr_colormap_energy_mode_unstable}(a), correspond to the frequencies of the most unstable eigenvectors of the Floquet analysis~\cite{CXYKT2021}. Thus, there is link of the presence of a weakly chaotic dynamics in the system, with the weakly unstable Floquet multipliers, namely the one with a small divergence from the unit circle~\cite{DRT1997,AB2006,CXYKT2021}.}

% 

%%%%%%%%% FIG %%%%%%%%%%%%
  \begin{figure}[t]
     \centering
     \includegraphics[width=0.7\columnwidth, height=1.\linewidth]{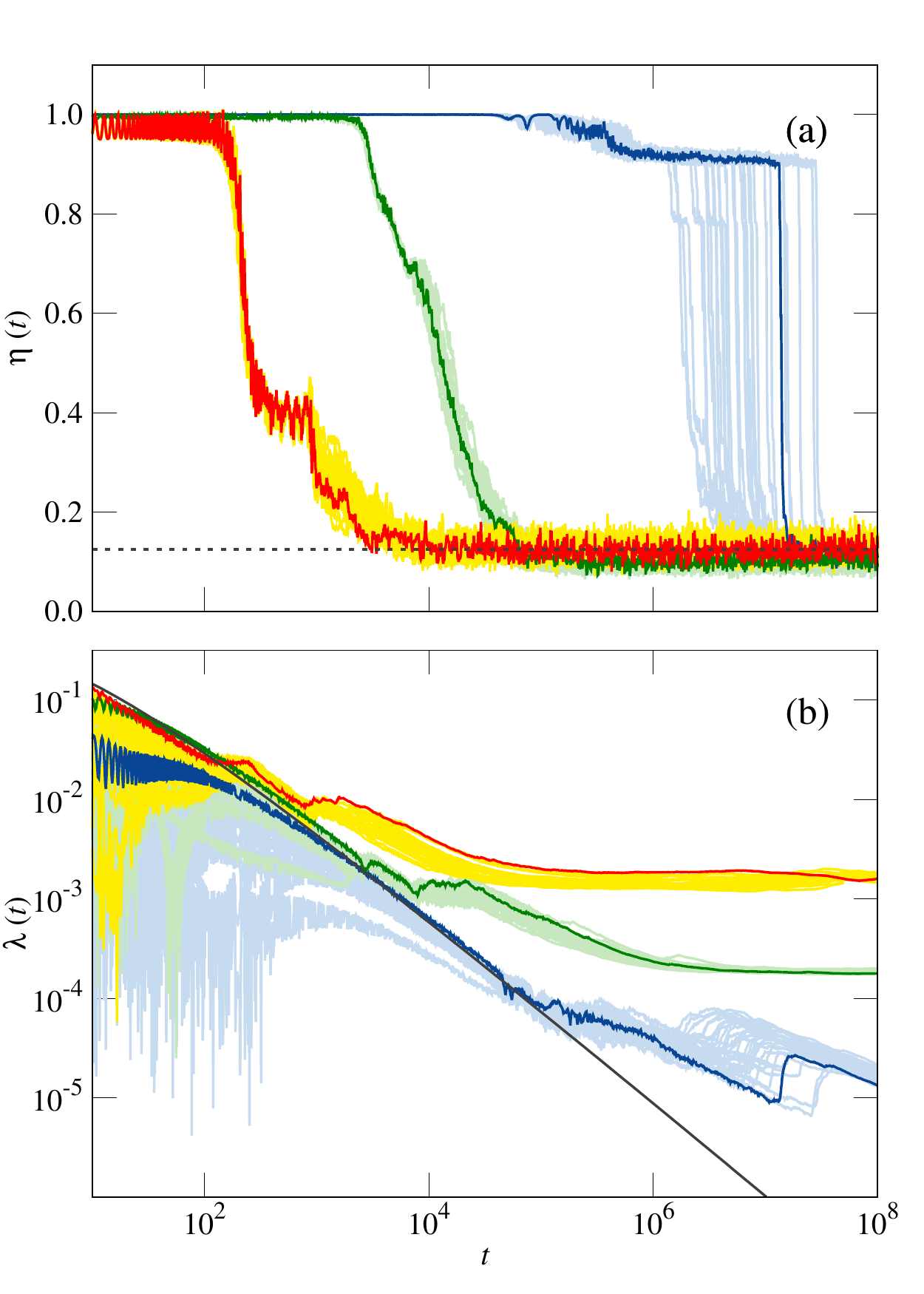}
     \caption{Temporal evolution of the (a) rescaled spectral entropy $\eta (t)$ [Eq.~\eqref{eq:resc_entr}] and (b) the ftMLE $\lambda(t)$ [Eq.~\eqref{eq:ftmle}] of the model [Eq.~\eqref{eq:hamilton_tkg}] for representative realizations  of the initial nonlinear topological edge states with $(\mathcal{H}=0.307, \Gamma=-0.8)$ [blue], $(\mathcal{H}=1.608, \Gamma=-0.8)$ [green] and $(\mathcal{H}=4.214, \Gamma=-0.8)$ [red] of Fig.~\ref{fig:distr_colormap_energy_mode_unstable}.
     The black dashed horizontal line in (a) measures $\langle \eta \rangle_G \approx 0.125$, the Gibbs average for the case $(\mathcal{H}=4.214, \Gamma=-0.8)$ [red curves in panels (a) and (b)].
     The shaded curves around the dark-colored curves' course represent the evolution of $30$ other realizations of perturbation of the same topological edge states.
     The black solid line in (b) shows the relation $\lambda(t)\propto\ln(t)/t$ observed in the case of regular motion.
    %  \color{red}{(i)Ammend this according to Fig.3(a) and edit the text accordingly.
    %  Redo Fig 3 and 4 with $\Gamma=0.8$ (generic or perturbed edge states).}
     }
     \label{fig:entr_mle_unstable}
 \end{figure}

On the other hand, in Figs.~\ref{fig:distr_colormap_energy_mode_unstable}(b,c), the normalized energy profile of the cases $(\mathcal{H}=1.608, \Gamma=-0.8)$ and $(\mathcal{H}=4.214, \Gamma=-0.8)$ are shown. These initial conditions correspond to strongly unstable nonlinear edge states with ``edge-bulk'' instability, which is caused by the collision of one \textit{bulk} mode (spatially extended) with an edge mode (localized) ~\cite{CXYKT2021}. This kind of instability is present even in the thermodynamic limit.
In contrast with the previous case of Fig.~\ref{fig:distr_colormap_energy_mode_unstable}(a),   %[Eq.~\eqref{eq:hamilton_tkg}], 
% \textcolor{blue}{(RC: I think the type of nonlinearity is changing here. So we can not attribute the following observation to the increase in energy only)} 
% more and more modes 
one observes the excitation of almost continuous bands (in the optical and acoustic frequencies) of modes. The size of these excited bands, grows increasing the value of the energy
% \textbf{GT: more and modes are excited or rather a continuous band of modes? I remember that it was the second, no? or maybe if one do it carefully to observe more and more and finally a continous region of modes to be excited?}
[see Figs.~\ref{fig:distr_colormap_energy_mode_unstable}(b) and (c)]. 
This efficient and broadband excitation %somehow resonace of almost all NMs 
may be related to the onset of the Chirikov criterion~\cite{IC1965}, which predicts an energy threshold above which effective (fast) energy transfer between NMs takes place~\cite{BI2005,LO2018}.
Since these energy transfers homogenize  chaos within the lattice interior, we also expect the chaotic dynamics to be stronger~\cite{DRT1997,AB2006,BI2005},
and the system to thermalize faster, as indeed it is observed in Figs.~\ref{fig:distr_colormap_energy_mode_unstable}(b) and (c).

\begin{figure}[t]
    \centering
    \includegraphics[width=\columnwidth, height=0.5\linewidth]{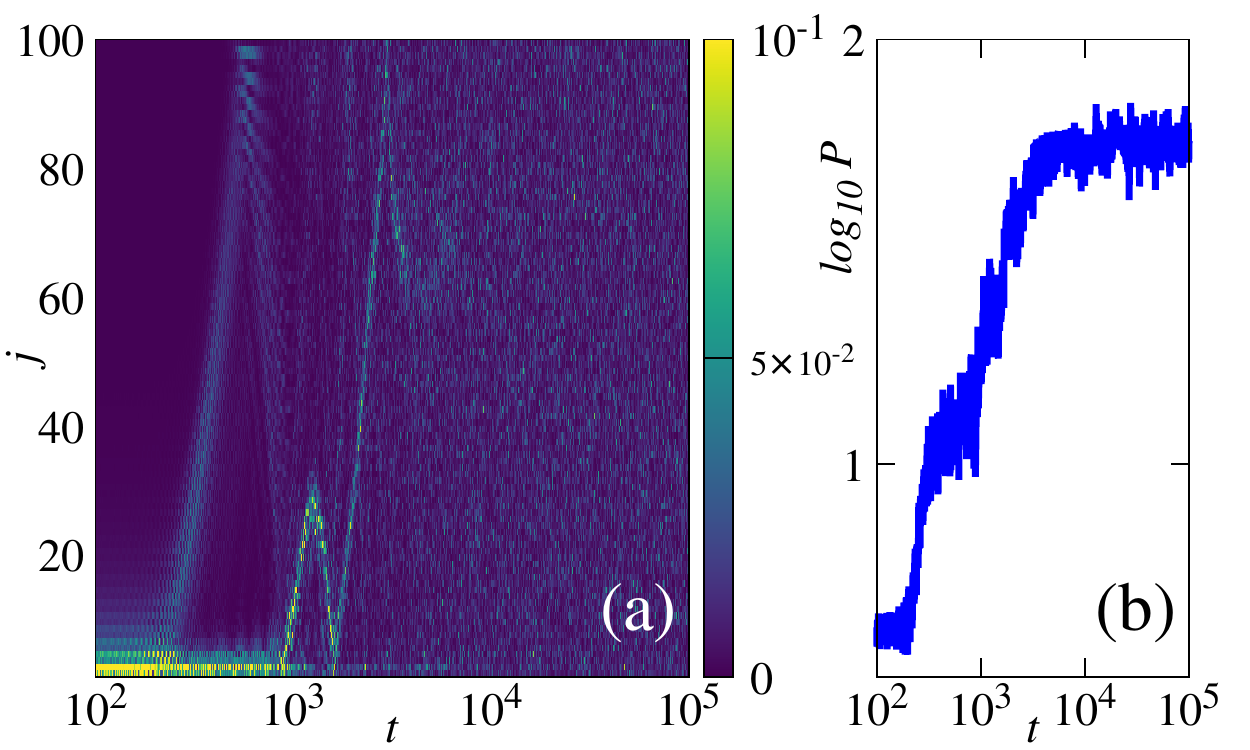}
    \caption{(a) Spatiotemporal profile of the normalized energy distribution $\xi_j (t)$ [Eq.~\eqref{eq:normalized_ener_per_site}] for the case $(\mathcal{H}=4.214, \Gamma=-0.8)$ of Figs.~\ref{fig:distr_colormap_energy_mode_unstable}(c) and~\ref{fig:entr_mle_unstable} (red curves in that figure).
    Each lattice site is colored according to the magnitude of its $\xi_j$ value [see colorscale at the right of (a)].
    (b) Evolution of the associated participation number $P(t)$ [Eq.~\eqref{eq:participation_number_ener_distr}].
    We see that the saturation of the entropy $\eta (t)$ [red curve in Fig.~\ref{fig:entr_mle_unstable}(a)] practically coincides with the one displayed by $P(t)$ in panel (b).
    % \textcolor{green}{HS: We need some additional ticks in the horizontal axes. Also the color bar needs more ticks.]}
    }
    \label{fig:colormap_energy_per_site_metastable_states}
\end{figure}

 \begin{figure*}[t]
     \centering
    \includegraphics[width=1\textwidth, height=0.275\linewidth]{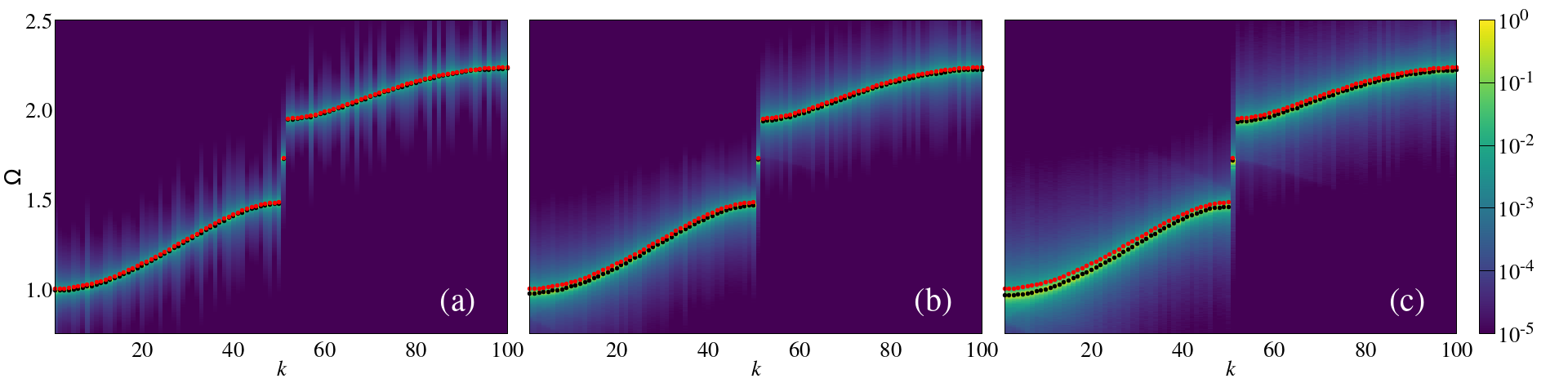}
     \caption{$\Omega -k$ dependence of the frequency shift distribution $\langle \lvert a_k \left(\Omega\right)\rvert^2\rangle$ for (a) $(\mathcal{H}=2.417, \Gamma=-0.8)$, (b) $(\mathcal{H}=4.409, \Gamma=-0.8)$ and (c) $(\mathcal{H}=7.019, \Gamma=-0.8)$.
      The red dots indicate the dispersion relation for the linearized system; whereas the black dots  indicate the renormalized dispersion relation for the nonlinear system  (see text for details).} 
     \label{fig:colormap_freq_shift_distribution}
 \end{figure*}
 
The thermalization and chaotization of  system [Eq.~\eqref{eq:hamilton_tkg}] are further confirmed from the computation of the rescaled entropy $\eta(t)$ [Eq.~\eqref{eq:resc_entr}], of the normalized energy per mode, and the ftMLE, $\lambda (t)$ [Eq.~\eqref{eq:ftmle}] (Fig.~\ref{fig:entr_mle_unstable}).
In Fig.~\ref{fig:entr_mle_unstable}(a), the rescaled spectral entropy $\eta$ [Eq.~\eqref{eq:resc_entr}] is plotted as a function of time $t$, for the three sets of parameters used in Fig.~\ref{fig:distr_colormap_energy_mode_unstable}.
All  curves start at $\eta = 1$, at time $t=0$ and then decline to settle toward values well approximated by the (Gibbs) ensemble average $\langle \eta \rangle_G$ [Eq.~\eqref{eq:thermodynamic_resc_entr}] with values $\approx 0.092$, $0.096$ and $0.125$ [black dashed line in Fig.~\ref{fig:entr_mle_unstable}(a)] for the cases with $\mathcal{H}=0.307$, $1.608$ and $4.214$ [blue, green and red curves in Fig.~\ref{fig:entr_mle_unstable}] respectively.
Since these relaxation processes are related to energy exchange between NMs, we expect to see a quantitative difference on how the transitions $\eta(t=0)=1\rightarrow \langle \eta \rangle_G$ are carried.

For the case with $(\mathcal{H}=0.307, \Gamma=-0.8)$ [blue curves in Fig.~\ref{fig:entr_mle_unstable}(a)], the relaxation of $\eta(t)$ takes longer, with a timid 
% \textbf{GT: not abrupt?}
decrease 
observed after the curve leaves the unity in the time interval from $\approx 10^5$ up to around a few $10^{7}$ units,
% between $10^{5}$ to $10^{6}$ 
which roughly corresponds to the interval on which we noticed the resonance of a small number of NMs responsible of the system's thermalization in Fig.~\ref{fig:distr_colormap_energy_mode_unstable}(a).
It is also interesting to relate this observation with the temporal behavior of the system's ftMLE, $\lambda$ [Eq.~\eqref{eq:ftmle}]. 
Indeed, in the time period mentioned above, we see that the computed $\lambda(t)$ of the system [blue curve in Fig.~\ref{fig:entr_mle_unstable}(b)] slightly diverges from the one of regular motion $\lambda(t)\propto\ln t /t $ [black solid line in Fig.~\ref{fig:entr_mle_unstable}(b)], confirming the presence of a weakly chaotic dynamics within the system.
As  time evolves further away from $t\approx10^7$, a saturation of $\eta (t)$ to values of the order of the Gibbs average $\langle \eta \rangle_G$ [Eq.~\eqref{eq:thermodynamic_resc_entr}] becomes evident.
This saturation of the $\eta (t)$, is also consistent with the level off of the $\lambda(t)$, which further deviates from the behavior observed for regular motion [$\lambda(t) \propto \ln t / t$], indicating the lattice's full chaotization, and equipartition.

On the other hand, for the cases with $(\mathcal{H}=1.608, \Gamma=-0.8)$ [green curves in Fig.~\ref{fig:entr_mle_unstable}] and $(\mathcal{H}=4.214, \Gamma =-0.8)$ [red curves in Fig.~\ref{fig:entr_mle_unstable}], the evolution of $\eta$ falls down quicker and the $\lambda (t)$ saturates faster toward larger numerical values than the $(\mathcal{H}=0.307, \Gamma=-0.8)$ case due to the faster energy spreading  and stronger chaotic behavior.
It is worth mentioning that in the case $(\mathcal{H}=4.214, \Gamma=-0.8)$, we observe the presence of an intermediate plateau in the evolution of $\eta (t)$ in Fig.~\ref{fig:entr_mle_unstable}(a), at $t\approx 10^3$, which is associated to the presence in the system of metastable states~\cite{DCF2017}
visible in Fig.~\ref{fig:colormap_energy_per_site_metastable_states}(a).
These transient states are unstable discrete breathers (DBs) originating from the degradation of the initial topological nonlinear edge state, chaotically scattering within the lattice and shedding their energy toward the bulk [Fig.~\ref{fig:colormap_energy_per_site_metastable_states}(a)].
In Fig.~\ref{fig:colormap_energy_per_site_metastable_states}(b), we compute the associated evolution of the participation number $P(t)$ [Eq.~\eqref{eq:participation_number_ener_distr}] and confirm that the disappearance of DBs coincides with the final saturation of the participation number for $t\approx 10^4$ time units.
% , hence thermal equilibrium.
% {\color{red}BMM:
% Can I mention somehow the Lyapunov time of the case of Fig. 5?
% }

% The nonlinear topological edges states, belongs to islands of quasi periodic orbits in the system's phase space whose size depends on the nonlinearity  (i.e. $$\mathcal{H}$) in which depending on the strength of the perturbations
% The shaded blue curves in Fig.~\ref{fig:entr_mle_unstable}

The numerical simulations depicted in this section clearly show the presence of two different routes to thermalization of our topological system, namely, in the weak and strong nonlinear (chaos) regimes.
In the weak nonlinear limit, the thermalization of the lattice is due to the activation of a few modes. This reminds what is expected from near exact resonance of NMs well describes in the framework of wave turbulence theory (see, e.g.~\cite{OVPL2015,LO2018,PCBLO2019}).
On the other hand, in the strong nonlinear regime, a large number of modes are initially resonating leading to stronger chaotic dynamics and faster thermalization. How this behavior is connected with the strong instability predicted by the linear Floquet stability analysis, and a potential Chirikov resonance-overlap conditions~\cite{IC1965,C1979}, is a subject under current investigation.

 \subsection{\label{subsec:frequency_renorm_&_Overalapping}Renormalized frequency and resonances broadening}

% {\color{red}BMM: Double check the phase you are imputing in your simulations.} 
In the following numerical calculations, we perform ensemble average (denoted by $\langle \cdot \rangle$) over $200$ to $600$ realizations, where different random phases $\varepsilon_k=\pm 10^{-2}$ are implemented in to [Eq.~\eqref{eq:init_state_localized}].
For these sets of initial conditions, we have checked that the computed average rescaled entropy $\langle \eta (t)\rangle$ matches with a good accuracy the one of thermalized modes in the Gibbs description $\langle \eta \rangle_{G}$ [Eq.~\eqref{eq:thermodynamic_resc_entr}] and that our average ftMLE, $\langle \lambda (t) \rangle$ has saturated to a constant positive value.
We then follow the same procedure as in Ref.~\cite{LO2018} in order to extract the spectral characteristics of our multidimensional nonlinear system [Eq.~\eqref{eq:hamilton_tkg}]. 
% \textcolor{green}{[HS: in order to do what?]}
That is to say, after reaching the Gibbs statistical equilibrium of the modal energies, we record the $a_k(t)$ [Eq.~\eqref{eq:normal_mode_variables}] for a time window of about $500T_1$ time units, where $T_1$ is the period of the mode with the smallest $\omega_k(\gamma)$.
Then for each mode $k$, the standard discrete Fourier transform~\cite{PTVF1996} is implemented, switching our independent variable $t \rightarrow \Omega$.
We generate  the ensemble average power spectrum $\langle \lvert a_k (\Omega) \rvert^2\rangle$, which has duly been rescaled by its maximum for each mode $k$.
Consequently, we derive the model's dispersion relation from our numerical data,
\begin{equation}
    \tilde{\omega}_k = \left\{\Omega/ \langle \lvert a_k (\Omega) \rvert^2\rangle = \max_{\Omega^\prime} \langle \lvert a_k \left(\Omega^\prime\right) \rvert^2\rangle\right\}.
    \label{eq:renorm_disper_relation}
\end{equation}
Note that if the system [Eq.~\eqref{eq:hamilton_tkg}] is linear, i.e. $\mathcal{H}=\mathcal{H}_2$, we have  $\langle \lvert a_k (\Omega) \rvert^2\rangle = \delta_{\Omega, \omega_k}$, and we get  $\tilde{\omega}_k=\omega_k$ from [Eq.~\eqref{eq:renorm_disper_relation}].
The $\Omega-k$ dependence of $\langle \lvert a_k (\Omega) \rvert^2\rangle$ is plotted for $(\mathcal{H}=2.417, \Gamma=-0.8)$ in Fig.~\ref{fig:colormap_freq_shift_distribution}(a), for $(\mathcal{H}=4.409, \Gamma=-0.8)$ in Fig.~\ref{fig:colormap_freq_shift_distribution}(b) and for $(\mathcal{H}=7.019, \Gamma=-0.8)$ in Fig.~\ref{fig:colormap_freq_shift_distribution}(c).
We observe that the calculated dispersion relation [black dots in Fig.~\ref{fig:colormap_freq_shift_distribution}] shifts away from the dispersion relation of the linearized system [red dots in Fig.~\ref{fig:colormap_freq_shift_distribution}] toward smaller values of the frequency, while at the same time a broadening of the distribution of frequency shift appears for each wave number.
This renormalization of the frequencies i.e., $\omega_k\rightarrow \tilde{\omega}_k$ 
% \textcolor{blue}{RC: I think we do not define ``renormalization'' anywhere. Is it getting the \textit{mean} of the frequency distribution for a specific $k$?}
, as well as the broadening of their distributions, becomes stronger, thus more evident for increasing values of $\mathcal{H}$ [Eq.~\eqref{eq:hamilton_tkg}]. 

In App.~\ref{app-sec:width_resonance}, we present additional results on the dependence of the renormalized frequency $\tilde{\omega}_k$ [Eq.~\eqref{eq:renorm_disper_relation}], the width of the frequency broadening $\chi_k$ [Eq.~\eqref{eq:width_freq_distr}] and the MLE $\Lambda$ [Eq.~\eqref{eq:mle}] on the energy (nonlinearity) of the system.

% It also seems more marked \textcolor{green}{HS: Not clear. What do you mean?]} for modes in the lower part of the spectrum, Fig.~\ref{fig:colormap_freq_shift_distribution}(c).
%It is worthwhile arguing the observations above in line with the recent works %in the field of lattice thermalization

The above observations are in line with the recent works on nonlinear lattice thermalization~\cite{GLC2005,JLZC2014,OVPL2015,LO2018,POC2018,PCBLO2019,PDLLO2021,FKS2009}. 
In those studies, the broadening of the width of the distribution of frequency shifts is attributed to the presence of the nonlinearity in the system, which introduces a degree of stochasticity (chaoticity) in the modes' interactions~\cite{IC1965,C1979}.
%  \textbf{GT: It is not very clear how figure 8 helps understanding this point.}
Besides, the nonlinear frequency shift was investigated for finite chain (e.g. $n=16,~32,~64$ and $128$) of various lattices including the monoatomic $\alpha$- and $\beta$-FPUT models and the KG systems as well as the diatomic $\alpha$-FPUT lattice.
In connection to the monoatomic $\beta$-FPUT, it was found that the shift in frequency is due to both trivial~\cite{GLC2005} and nontrivial~\cite{JLZC2014} four-wave resonances, while for the $\alpha$-FPUT counterpart, only trivial four-wave resonances are the drivers of this shift~\cite{OVPL2015}.
The later type of mode resonance is also relevant for the quartic KG system~\cite{S2011,POC2018}.
Furthermore, in the case of the diatomic $\alpha$-FPUT chain, three-wave resonances are responsible for the renormalization of frequencies~\cite{PDLLO2021}.
% Overall, each of these works pointing to different types of resonnances at the origin of the lattice thermalization in the weak nonlinear limit.
As such, it is pertinent to give a closer look toward which resonant processes are involved in the present model of Fig~\ref{fig:description_ssh_kg}, something we plan to tackle in future endeavours.

\subsection{\label{subsec:effective_dispersion}Symmetry of the renormalized squared dispersion relation}

\begin{figure}[t]
    \centering
    \includegraphics[width=0.75\columnwidth]{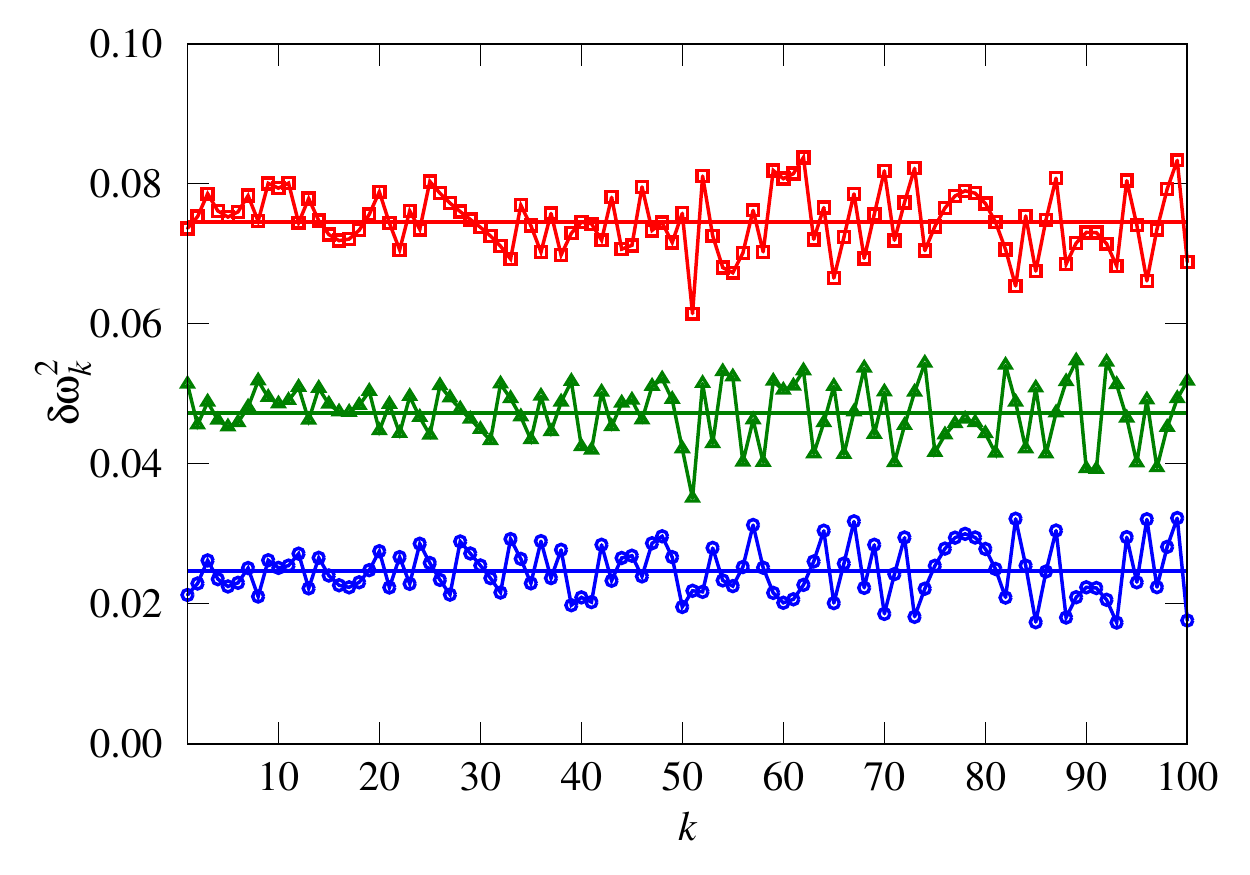}
    \caption{Change in the squared frequency $\delta \omega_k^2 = \omega_k^2 - \tilde{\omega}_k^2$ [Eq.~\eqref{eq:change_in_freq}] of the dispersion relation at different energy levels: $\mathcal{H}=2.417$ (blue line-connected circles), $\mathcal{H}=4.214$ (green line-connected triangles) and $\mathcal{H}=7.019$ (red line-connected squares).
    For all cases, $\Gamma=-0.8$.
    The horizontal solid lines of the same colors indicate the average values $\overline{\delta \omega^2}$ of the relevant data points they fit:
    $\overline{\delta \omega^2} \approx 0.025\pm0.004$, $0.047\pm0.004$ and $0.074\pm 0.004$, respectively for the blue, green and red line-connected  points.
    } 
    \label{fig:change_in_freq}
\end{figure}

We now look on potential symmetry properties of the numerically computed squared renormalized dispersion relation, $\tilde{\omega}_k^2$ [Eq.~\eqref{eq:renorm_disper_relation}].
As we stressed in Sec.~\ref{sec:topological_edge_states}, an important characteristic of the eigenvalues  $\omega_k^2$ is their symmetry about their mid-gap, a consequence of chiral symmetry of the dynamical matrix in the linear limit i.e.~$\Gamma=0$ in $\mathcal{H}$ in Eq.~\eqref{eq:hamilton_tkg} (see also Fig~\ref{fig:linear_disper_rel_&_topological_states_spatial_profile}).
It is therefore relevant to investigate what is the effect of nonlinearity on this symmetry, by checking the renormalized dispersion relation.
In Fig.~\ref{fig:change_in_freq}, we plot the change in the squared frequency
\begin{equation}
    \delta\omega_k^2 = \omega_k^2 - \tilde{\omega}_k^2= \left(1 - \eta_k^2 \right)\omega_k^2, \quad    \eta_k = \frac{\tilde{\omega}_k}{\omega_k},
    \label{eq:change_in_freq}
\end{equation}
for each wave number, $k$ using three set of parameters $(\mathcal{H}=2.417, \Gamma=-0.8)$ (blue circles), $(\mathcal{H}=2.417, \Gamma=-0.8)$ (green triangles) and $(\mathcal{H}=7.019, \Gamma=-0.8)$ (red squares).
In Eq.~\eqref{eq:change_in_freq} $\eta_k$ is the renormalization factor of the frequency of mode $k$~\cite{JLZC2014}.
Note that $\eta_k=1$ in the linear limit, while being $\eta_k<1$ for $\tilde{\omega}_k<\omega_k$ [see e.g., Figs.~\ref{fig:frequency_shift_&_width_vs_energy}(a), (c) and (e) in App.~\ref{app-sec:width_resonance}] in case $\Gamma<0$~\cite{LO2018}.
On the other hand, if $\delta \omega_k^2$ remains practically constant for all values of $k$, we can conclude that the squared renormalized dispersion relation is uniformly shifting away from the linear one, keeping thus its subsequent symmetry around the renormalized, squared mid gap frequency $\tilde{\omega}_{k=51}^2$.
Remarkably,  in Fig.~\ref{fig:change_in_freq}, we see that this is the case for all the studied energy values.
We note that increasing the value of $\mathcal{H}$ results in higher  $\delta \omega_k^2$ values which are fluctuating around the averages $\overline{\delta \omega^2} \approx 0.025$, $0.047$ and $0.074$, blue, green and red horizontal lines, for respectively  $\mathcal{H}=2.417$, $4.214$ and $7.019$.
Hence, we deduce that the contribution of the nonlinear part of the system renormalizes the squared dispersion relation $\omega_k^2$ in a way that it retains the symmetry 
inherited from chiral symmetry of the dynamical matrix of the linearized model.
% This is an interesting results which deserves more attention in the future.

\begin{figure*}[t]
     \centering
     \includegraphics[width=1\textwidth, height=0.275\linewidth]{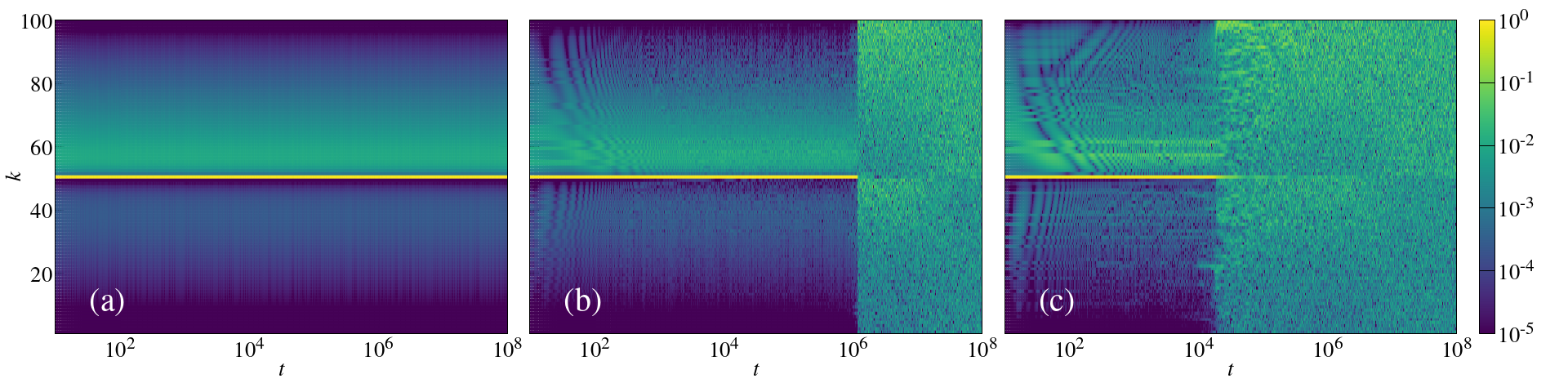}
     \caption{Similar to Fig.~\ref{fig:distr_colormap_energy_mode_unstable} but for three different perturbation strengths $\varepsilon$ [Eq.~\eqref{eq:init_state_localized}] of the nonlinear topological edge state at ($\mathcal{H}$=3.779, $\Gamma$=0.8) with (a) $\varepsilon=0.01$, (b) $\varepsilon=0.25$ and (c) $\varepsilon=1.0$, see text for details.
     Each mode's number is colored according to the intensity of its energy.
    %  The color box at the left is in logarithmic scale.
     }
     \label{fig:distr_energy_mode_regular_cases}
 \end{figure*}
 
%%%%%%%%%%%%%%%%%%%%%%%%%%%%%%%%%%%%%%%%%%%%%%%%%%%%%%%%%%
\subsection{\label{subsec:neigborhood_of_stable_states}Neighborhood of stable nonlinear topological edge states}

% \textcolor{green}{[HS: We have to discuss what we will do with this section. I have several comments but I will implement them after finalizing the content of this section.  ]}
% {\color{red} 
% \begin{itemize}
%     \item Check the statistical characteristics of thermalization of the previous section for cases which thermalized here. for example, take a single breather with a perturbation which lead to $100\%$ chaos and do the statistics (construct all the curves of the previous sections).
%     % \item Change the threshold of for thermalization $\eta_{th}$ to a value closer to Gibbs thermal equilibrium $0.2$.
%     % \item Rewrite this section focussing on the thermalization. Can mention that similar trends is observed for chaos.
%     % \item Redo the figure of the fraction for only the thermalization (with the new threshold and another one with the MLE and entropy.
% \end{itemize}
% }

% \textbf{GT:Up to now, we have considered the cases of linearly unstable topological edge states. One of the remarkable results of~\cite{CXYKT2021}, is the existence of linearly stable topological edge states under strong nonlinearity.}

Up to now, we have considered the cases of linearly unstable topological edge states. One of the remarkable results of Ref.~\cite{CXYKT2021}, is the existence of linearly stable topological edge states under strong nonlinearity.
Here, we want to understand how robust the stable nonlinear topological edge states are under the system's perturbation. 
We consider the nonlinear topological edge state at $(\mathcal{H}=3.779, \Gamma=0.8)$ [shown in Fig.\ref{fig:linear_disper_rel_&_topological_states_spatial_profile}(c)] applying random perturbations of the form $\varepsilon_k = \pm \varepsilon$ with $\varepsilon>0$ referred to as perturbation parameter, see Fig.~\ref{fig:distr_energy_mode_regular_cases}.
The amplitude profiles of $\nu_k(t)$ corresponding to representative realizations of three different perturbation parameters [see Eq.~\eqref{eq:init_state_localized}] with $\varepsilon=0.01$, Fig.~\ref{fig:distr_energy_mode_regular_cases}(a), $\varepsilon=0.25$, Fig.~\ref{fig:distr_energy_mode_regular_cases}(b) and $\varepsilon=1$ in  Fig.~\ref{fig:distr_energy_mode_regular_cases}(c) are plotted.
For the smallest perturbation parameter [Fig.~\ref{fig:distr_energy_mode_regular_cases}(a)], the initially excited modes $a_k^{b}(t=0)$ retain their energy $E_k = \omega_k \lvert a_k (t=0) \rvert^2$ and the lattice remains unthermalized up to the largest simulation time $t=10^8$. 
Thus, in this case, the nonlinear topological state appears to be stable under the added perturbations.
As the magnitude of the perturbation parameter $\varepsilon \rightarrow \mathcal{O}(1)$ [Figs.~\ref{fig:distr_energy_mode_regular_cases}(c) and (d)], we see that the obtained initial states eventually lead to equipartition in the lattice after time scales that vary with $\varepsilon$ [Eq.~\eqref{eq:init_state_localized}].
This hints us on the fact that there exists a limiting value of the perturbation $\varepsilon_c$ above which the coherence of the stable nonlinear topological edge state is lost.

\begin{figure}[!]
    \centering
    \includegraphics[width=0.95\columnwidth, height=0.8\linewidth]{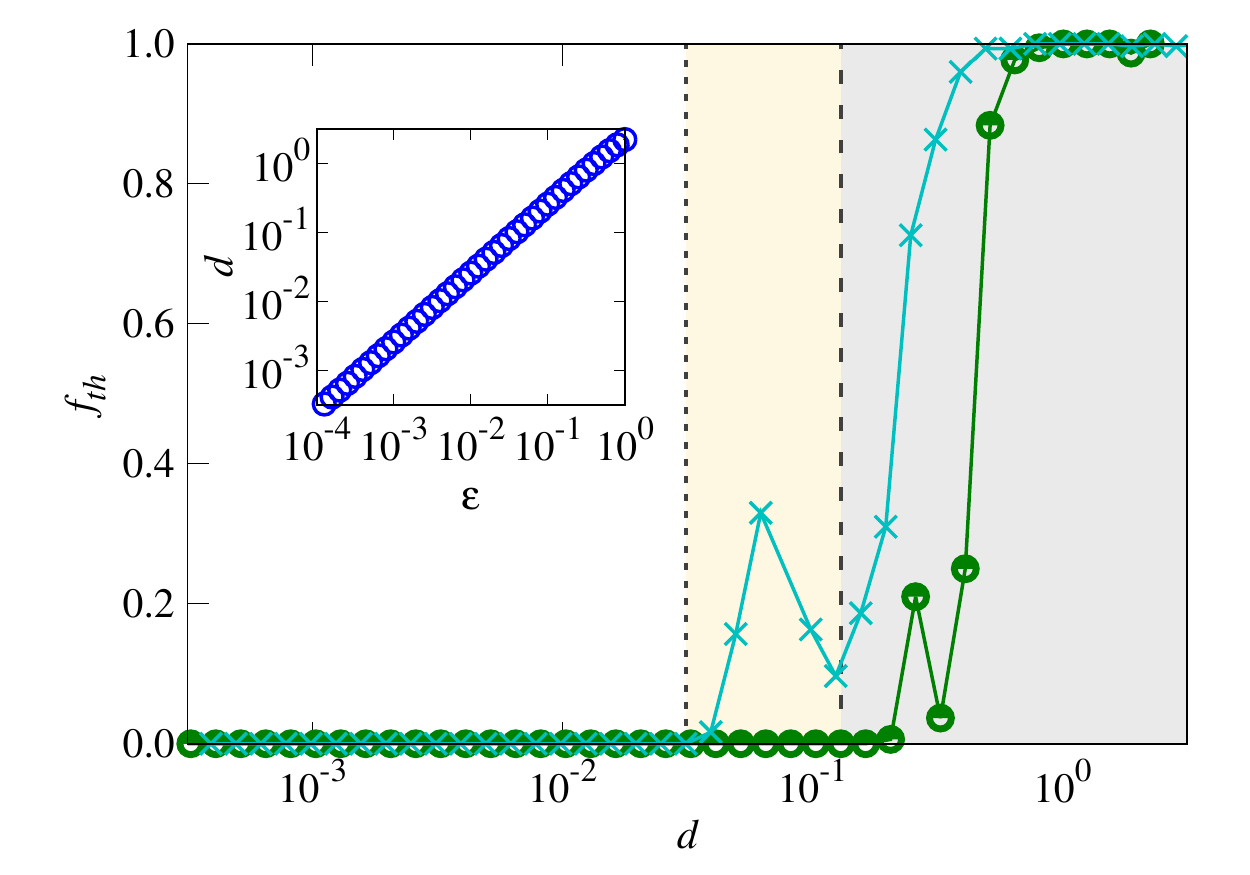}
    \caption{Fraction of thermalized $f_{th}$ [Eq.~\eqref{eq:fraction_of_chaos_therm}] states of the system [Eq.~\eqref{eq:hamilton_tkg}] at time $t=10^{8}$ as function of the distance $d$ [Eq.~\eqref{eq:size_regular_island}] from the linearly stable nonlinear topological state in the phase space with $(\mathcal{H}=3.779, \Gamma=0.8)$ [green line-connected circles] and $(\mathcal{H}=8.367, \Gamma=0.8)$ [cyan line-connected crosses].
    The inset panel shows the dependence of $d$ against the perturbation parameter $\varepsilon$ of Eq.~\eqref{eq:init_state_localized}.
    The dashed vertical line indicates $d_c\approx 0.13$ and the dotted one $d_c\approx 0.031$ (see text for details).
    }
    \label{fig:fraction_of_chaotic_trajectories}
\end{figure}

In general, the stability properties of autonomous Hamiltonian system's are well described in the framework of phase space dynamics.
Indeed, we expect stable nonlinear topological edge states to belong to regular `islands' surrounded by chaotic `sea'~\cite{AB2006,BKOMS2020}.
In the prospective of Secs.~\ref{subsec:charact_unstable_topo_states} and~\ref{subsec:frequency_renorm_&_Overalapping}, a system's orbit belonging to a regular island does not lead to the system's thermalization in real (NM) space, while the one within the chaotic sea does.
If we further assume that the stable topological edge state lies at the center of the island of stability, we define the distance 
\begin{equation}
    d=\lVert \boldsymbol{X}_{\varepsilon}(0)-\boldsymbol{X}^{b}(0)\rVert,
    \label{eq:size_regular_island}
\end{equation}
which tells us how far we are from the topological edge state, i.e., the radius of the regular island~\cite{BKOMS2020}.
In Eq.~\eqref{eq:size_regular_island}, $\boldsymbol{X}^{b}$ is the topological edge state obtained from the procedure explained in Sec.~\ref{sec:topological_edge_states} and $ \boldsymbol{X}_{\varepsilon}$ its perturbed state [which depends on $\varepsilon$ of Eq.~\eqref{eq:init_state_localized}], while $\lVert \cdot \rVert$ stands again for the Euclidean norm.
In order to obtain a reliable measurement of the phase space observable $d$ [Eq.~\eqref{eq:size_regular_island}], the phase space landscape has to remain unchanged as we perturb $\boldsymbol{X}^{b}$.
Consequently, after applying Eq.~\eqref{eq:init_state_localized}, we particularly take care that the final initial condition $\boldsymbol{X}_{\varepsilon}(0)$ possesses the same energy as $\boldsymbol{X}^{b}$ up to the $10$th digit via a Newton-Raphson procedure~\cite{PTVF1996}.
The numerically computed values of $d$ show that it grows exponentially with increasing $\varepsilon$ values [see inset of Fig.~\ref{fig:fraction_of_chaotic_trajectories}].

% {\color{red}BMM: I am here.}

For a specific perturbation strength $\varepsilon$, the choice of different sets of phases $\varepsilon_k =\pm \varepsilon$, gives initial conditions of distance $d$ from the topological edge state in the phase space.
We categorize each of these initial conditions as leading or not to thermalization by computing the rescaled spectral entropy $\eta(t)$~[Eq.~\eqref{eq:spectral_entr}].
Then, based on observations of our numerical simulations like that of thermalized states which have $\eta(t)\approx \langle \eta \rangle_G$, we introduce the threshold
\begin{equation}
    % \lambda_{ch} = 10^{-6}, \quad 
    \eta _{th} = 0.2,
    \label{eq:chaos_therm_thresholds}
\end{equation}
such that trajectories with
% $\lambda(t=10^{8}) \ge \lambda_{ch}$ or 
$\eta (t=10^{8}) \leq \eta_{th}$ are considered to
% be chaotic or that they would 
lead to the lattice thermal equilibrium.
In this way, we can define the fraction~\cite{MAPS2009,M2014}
\begin{equation}
    % f_{ch}=\frac{R_{ch}}{R}, \quad
    f_{th}=\frac{R_{th}}{R},
    \label{eq:fraction_of_chaos_therm}
\end{equation}
of 
% chaotic orbits and 
orbits conducing to the lattice's thermalization.
Here, 
% $R_{ch}$ is the number of chaotic trajectories, 
$R_{th}$ is the number of orbits leading to energy equipartition and $100 \leq R \leq 300$ is the total number of initial conditions of distance $d$ from the topological edge state $\boldsymbol{X}^{b}(t=0)$.
 
To explore the neighborhood of the linearly stable, nonlinear, topological edge states, we calculate the fraction of 
% chaos and 
thermalized states, 
% $f_{ch}$ and
$f_{th}$ [Eq.~\eqref{eq:fraction_of_chaos_therm}] at different distances, $d$ [Eq.~\eqref{eq:size_regular_island}] from
the topological, nonlinear edge state in the phase space for the parameters $(\mathcal{H}=3.779, \Gamma =0.8)$ and $(\mathcal{H}=8.367, \Gamma=0.8)$, Fig.~\ref{fig:fraction_of_chaotic_trajectories}.
We see that for $d\rightarrow 0$ all the system orbits are regular i.e., 
% $f_{ch}\rightarrow 0$ and 
$f_{th}\rightarrow 0$.
On the other hand, for $d\rightarrow \mathcal{O}(1)$, all the orbits lead to the system thermalization as 
% $f_{ch}\rightarrow 1$ and 
$f_{th}\rightarrow 1$.
% It is worth remarking that in the case $d=\mathcal{O}(1)$, the initial condition [Eq.~\eqref{eq:init_state_localized}] becomes similar to the one of the `generic' initialization of lattice models seen in Refs.~\cite{BCP2013,JLZC2014,OVPL2015,LO2018,PCBLO2019,PDLLO2021} which is aimed to lead to a faster thermalization of the system. Thus its trajectory needs to belong to the chaotic sea in the system phase space from the beginning of the evolution. 

For intermediate values of $d$, the transition of $f_{th}=0\rightarrow 1$ takes place for all nonlinear strengths.
% It is remarkable to observe that for both parameters $(\mathcal{H}=0.3779, \Gamma=0.8)$, and $(\mathcal{H}=0.837, \Gamma=0.8)$ the fraction of chaos $f_{ch}$ [blue and red circle curves] are always larger than the fraction of thermalized states as one expect $f_{ch}$ to be a limiting case of $f_{th}$: the system must chaotize/randomize first before thermalizing~\cite{MAPS2009,SGF2013}. 
% \textbf{GT: You are the experts here, but to me this point (chaoticity cs thermalization) is quite technical. Up to know we did not make such distinction. We talked about Chaos and the resulting thermalization, while here we want to make this extra distinction. It can be chaotic but not thermalized right? and for the time scales that we are talking. Is it possible to have chaos and the nonlinear lattice would not thermalized for example? If yes, i see this statement and analysis here. In addition, looking the figure I do not see big differences and probably these small differences are due to the arbitrary definitions of the thresholds. Personally I prefer to keep only the thermalized threshold.}
The critical distance $d_c$ above which $ f_{th}>0$ represents a length scale above which the system is likely to thermalize and the topological edge state to loose its robustness to the perturbation.
Interestingly enough, we observe that $d_c$ decrease with increasing energy as $d_c\approx 0.13$ [dashed vertical line in Fig.~\ref{fig:fraction_of_chaotic_trajectories}] for the case with the smallest energy $(\mathcal{H}=3.779, \Gamma=0.8)$, dotted curves of Fig.~\ref{fig:fraction_of_chaotic_trajectories} and $d_c\approx 0.031$ [dotted dotted line in Fig.~\ref{fig:fraction_of_chaotic_trajectories}] for the case with the largest energy  $(\mathcal{H}=8.367, \Gamma=0.8)$, crossed curves of Fig.~\ref{fig:fraction_of_chaotic_trajectories}.
Once again, this observation is in agreement with the fact that islands of stability tend to disappear for large energy (nonlinearity) values in conservative Hamiltonian models~\cite{HH1964}. 

% One of the advantage of the phase space approach to characterize the robustness of stable topological state stands on the fact that $d$ is a distance in a multidimensional phase space which is ultimately independent of the types of initial conditions.
% Consequently a single $d$ value encompasses any types of perturbations $d$ distant from the stable topological edge state.

\begin{figure}[t]
    \centering
    \includegraphics[width=\columnwidth, height=\linewidth]{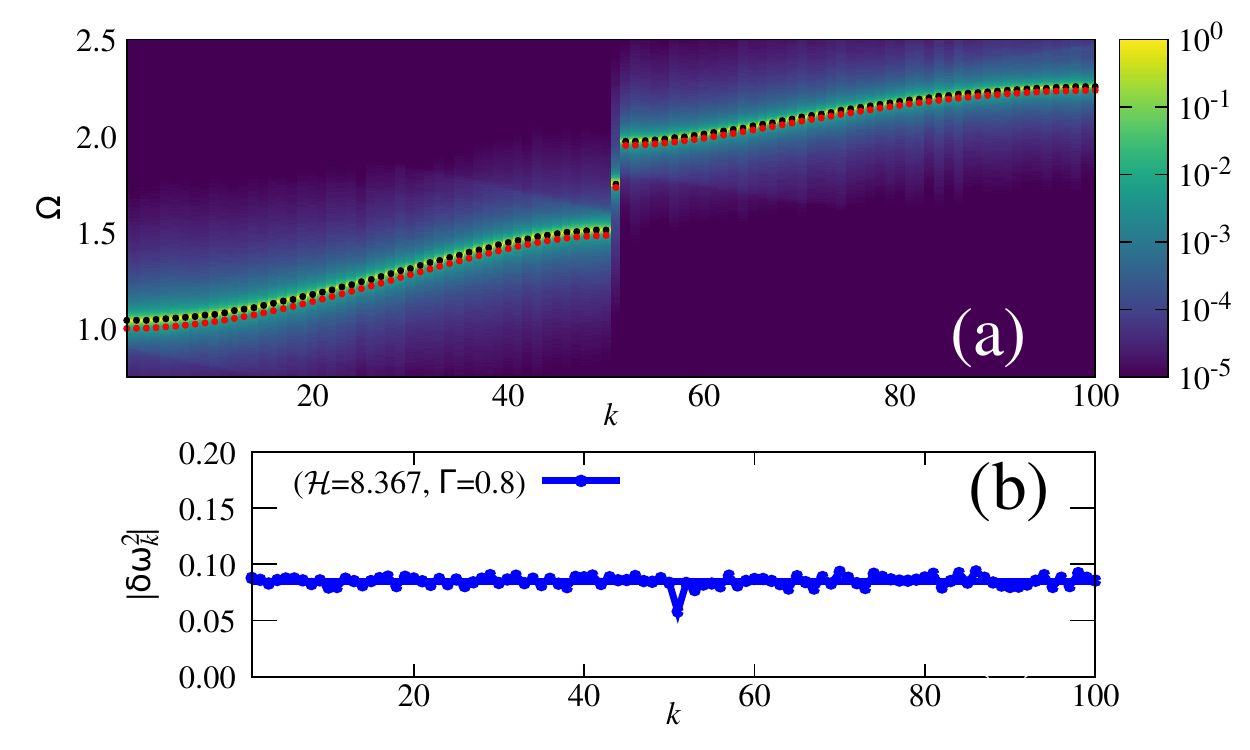}
    \caption{
        (a) Similar to Fig.~\ref{fig:colormap_freq_shift_distribution} but for a perturbation of the stable topological state with $(\mathcal{H}=8.367, \Gamma=0.8)$ at distance $d\approx0.285$ (see text for details).
        The red dotted line represents the dispersion relation of the linearized system $\omega_k$, while the black dotted line is the calculated dispersion relation $\tilde{\omega}_k$ [Eq.~\eqref{eq:renorm_disper_relation}].
        (b) Change in the squared frequency $\lvert \delta \omega_k^2\rvert$ [Eq.~\eqref{eq:change_in_freq}] (blue line-connected points) of the dispersion relations in panel (a).
        The value of $\lvert \delta \omega_k^2\rvert$ is oscillating around the average $ \overline{\delta \omega^2}=0.085 \pm 0.005$ [blue bold line in (b)].
    }
    \label{fig:colormap_stiff_perturb_01}
\end{figure}

Eventually when a stable topological edge state destabilizes due to the addition of perturbations, another related question is to know whether the resulting thermalized lattice state possesses similar spectral properties as for the unstable topological edge states seen in Secs.~\ref{subsec:charact_unstable_topo_states} and~\ref{subsec:frequency_renorm_&_Overalapping}. 
The noticeable difference between stable and unstable nonlinear topological edge states leans on the fact that their existence depends on the type of nonlinearity within the lattice model~\cite{CXYKT2021}.
Therefore, stable topological edge states are observed for stiffening nonlinear coefficients, i.e. $\Gamma>0$ in $\mathcal{H}$ [Eq.~\eqref{eq:hamilton_tkg}], while unstable topological edge states
% investigated in Secs.~\ref{subsec:charact_unstable_topo_states} and~\ref{subsec:frequency_renorm_&_Overalapping} 
are mostly associated with softening nonlinear strengths, i.e. $\Gamma <0$ in $\mathcal{H}$ [Eq.~\eqref{eq:hamilton_tkg}].

We numerically investigate the thermalization of the lattice [Eq.~\eqref{eq:hamilton_tkg}], when initially small random phases $\varepsilon_k=\pm 10^{-2}$ [Eq.~\eqref{eq:init_state_localized}] are added to the modes of an unstable perturbation at distance $d\approx 0.285$ of the stable topological edge state with parameter $\left(\mathcal{H}=8.367, \Gamma=0.8\right)$.
We then follow the same procedure describe in Sec.~\ref{subsec:frequency_renorm_&_Overalapping} in order to extract the spectral properties of the system at thermal equilibrium.
In Fig.~\ref{fig:colormap_stiff_perturb_01}(a), we plot the density spectrum $\langle \left\lvert a_k \left(\Omega\right)\right\rvert^2\rangle$ in the $\Omega-k$ space and extract the system's dispersion relation $\tilde{\omega}_k$ [Eq.~\eqref{eq:renorm_disper_relation}].
These results show that the henceforth calculated dispersion relation $\tilde{\omega}_k$ [black points in Fig.~\ref{fig:colormap_stiff_perturb_01}(a)] is shifted upward compared to the one of the linearized model $\omega_k$ [red points in Fig.~\ref{fig:colormap_stiff_perturb_01}(a)].
On the other hand, each mode, $k$, is characterized by a distribution of frequencies as seen in Sec.~\ref{subsec:frequency_renorm_&_Overalapping}, which makes it possible, also, in the stiffening case ($\Gamma>0$) for wave-wave interactions.
In addition, we plot in Fig.~\ref{fig:colormap_stiff_perturb_01}(b) the change in the squared frequencies $\lvert \delta \omega^2 _k\rvert$ [Eq.~\eqref{eq:change_in_freq}].
We find that, for increasing value of the wave number, $k$, the $\lvert \delta \omega_k^2\rvert$ remains practically constant, oscillating around the average value $\overline{\delta \omega^2}\approx 0.085$. 
For this reason, we also conclude that the lattice's thermal equilibrium in the case of stiffening nonlinearity is characterized by a renormalized dispersion relation which preserves the chiral symmetry of the underpinning system's dynamical matrix.
% the case of stiffening nonlinearity \ldots.

% Thus the model's thermal equilibrium in case of stiffening nonlinearity is also characterized by a renormalized dispersion relation which preserve the chiral symmetry of the system's dynamical matrix, Fig.~\ref{fig:colormap_stiff_perturb_01}.
% That being said, the change in the squared frequencies $\lvert \delta \omega^2 _k\rvert$~\eqref{eq:change_in_freq} is practically constant for all modes, as what we observe is the meandering of the former around an average value $\lvert\delta \omega^2\rvert\approx xxxx$.
% for an initial state at distance $d\approx0.285$ from the topological state with parameter $\left(\mathcal{H}=8.367, \Gamma=0.8\right)$.
% Once again, for the ensemble averaging, random phases $\varepsilon_k=\pm 10^{-2}$ of the mode are applied in the former state following~\eqref{eq:init_state_localized}.

%%%%%%%%%%%%%%%%%%%%%%%%%%%%%%%%%%%%%%%%%%%%%%%%%%%%%%%%%
\section{\label{sec:conclusion}Conclusion}

We conducted a detailed numerical study of the energy spreading in a nonlinear topological lattice. In particular, we focused on the long-time dynamics of the nonlinear edge states, which are obtained by nonlinear continuation of the edge states of the linearized lattice. Linearly \textit{unstable} edge states were shown to be delocalized from the edges in Ref.~\cite{CXYKT2021}. Here we showed that such delocalization is followed by chaos and leads to the thermalization of the lattice. The time to thermalization reduced with the increase of the nonlinear edge state's energy. We also observed an effective renormalization of the  dispersion relation of the linearized model, and a broadening of the distribution of the renormalized frequencies for all modes within the lattice.   

We also investigated the robustness of the linearly \textit{stable} nonlinear topological states by adding random perturbations to the initial state, and thus moving away from its trajectory in the phase space.
Our results showed that these nonlinear edge states remain robust up until a threshold of perturbation, beyond which we witness the loss of robustness because the related trajectory entering into a chaotic region of the phase space. Also, this threshold decreases upon increasing the energy of the system. We also observed an effective renormalization of the  dispersion relation similar to the case above. 

Finally, we discovered that the effective renormalization of the dispersion relation retains a unique symmetry, i.e, its square is symmetric about the squared mid-band frequency, reminiscent of chiral symmetry in the linearized system. This indicates that the classical symmetries of linear topological lattices in general could have their signature in the renormalized dispersion relation of their nonlinear counterparts. 

We believe that this work provides an interesting outlook into the thermalization of topological lattice systems and especially starting with initial conditions in the neighborhood of unstable nonlinear edge states.
These states are (unstable) quasi-periodic orbits in the phase space, whose evolutions usually transit smoothly from regular to chaotic behaviors, therefore providing us with the opportunity to better understand the causes of energy spreading in nonlinear discrete lattices.
Nonetheless, these results raised a number of significant questions worth investigating in the future.
More specifically, how the Floquet stability of the topological edge mode~\cite{CXYKT2021} connects to the mode's resonance theory (e.g., discrete mode resonances~\cite{OVPL2015}, Chirikov resonance-overlap~\cite{C1979}) of thermalization of discrete lattice model?
As such, expressing the mode resonance conditions of the present system along the directions of~\cite{OVPL2015,LO2018,PCBLO2019,PDLLO2021} is relevant for this problem.
Additionally, understanding how the above mechanisms change with the  type of initial conditions and/or the nonlinearity of the system is also an interesting line in order to construct a more general framework of thermalization on topological lattice systems.
Such work could give us a ground to better understand the transport of energy in nonlinear topological systems and comparing them with those of
generics lattice models (see e.g.~\cite{CBTD2016}).
Thus, appreciate the influence of topologically protected nonlinear modes on the long time dynamics of topological systems.
At last, extending this work to lattices with higher spatial dimensions is also a natural extension of the present study.

\begin{acknowledgments}
The authors would like to thank the Centre for High Performance Computing (CHPC) of South Africa~\cite{chpconline} as well as the High Performance Computing facility of the University of Cape Town~\cite{hpcuctonline} for providing their computational resources.
\end{acknowledgments}

%%%%%%%%%%%%%%%%%%%%%%%%%%%%%%%%%%%%%%%%%%%%%%%%%%%%%%%%%
\appendix

%%%%%%%%%%%%%%%%%%%%%%%%%%%%%%%%%%%%%%%%%%%%%%%%%%%%%%%%%
\section{\label{app-sec:symplectic_integration}Symplectic integration of the equations of motion and variational equations of the Su--Schrieffer--Heeger type Klein-Gordon lattice model}
In this section we present how we integrate the equations of motion and the variational equations of the Su--Schrieffer-Heeger type Klein-Gordon lattice model [Eq.~\eqref{eq:hamilton_tkg}] using a symplectic integration scheme along with the {\it tangent map method}~\cite{SG2010,GS2011,GES2012}.
The Hamiltonian $\mathcal{H}$ [Eq.~\eqref{eq:hamilton_tkg}] can be separated into two integrable parts, namely
\begin{equation}
    \mathcal{H} = \mathcal{A}\left(\boldsymbol{p}\right) + \mathcal{B}\left(\boldsymbol{x}\right),
    \label{eq:splitting_ham_1}
\end{equation}
where $\boldsymbol{x} = \left(x_1, x_2, \ldots, x_n\right)$ and $\boldsymbol{p} = \left(p_1, p_2, \ldots, p_n\right)$ are respectively the system's conjugate position and momentum vectors in the phase space, which is characterized by a vector $\boldsymbol{X}(t) = \left(\boldsymbol{x}(t), \boldsymbol{p}(t)\right)$.
In this context, we have 
% [HS: Similar comment as for equation (1).]
\begin{equation}
    \mathcal{A}\left(\boldsymbol{p}\right) = \frac{1}{2}\sum _{j=1}^{n} p_j^2,
    \label{eq:ham_a}
\end{equation}
and
\begin{equation}
    \begin{split}
        &\mathcal{B}\left(\boldsymbol{x}\right) =   \sum_{j=1}^{n} \left[\frac{\gamma_0}{2} x_j^2 + \frac{\Gamma}{4}x_j^4\right] \\
        & + \frac{1}{4} \sum_{\substack{j=1 \\ j = \text{odd}}}^{n} \left[(1 + \gamma) \left(x_j - x_{j - 1}\right)^2 + (1 - \gamma)\left(x_{j + 1} - x_j\right)^2 \right] \\
        & + \frac{1}{4} \sum_{\substack{j=0 \\ j = \text{even}}}^{n} \left[ (1 - \gamma) \left(x_j - x_{j - 1}\right)^2 + (1 + \gamma)\left(x_{j + 1} - x_j\right)^2 \right] 
    \end{split}.
    \label{eq:ham_b}
\end{equation}
where $n$ is the number of nonlinear oscillators.
Note that in both $\mathcal{A}$ [Eq.~\eqref{eq:ham_a}] and $\mathcal{B}$ [Eq.~\eqref{eq:ham_b}] fixed and free boundary conditions are respectively applied at the left ($j=0$) and right ($j=n+1$) edges of the lattice.

In the Lie formalism, the Hamilton equations of motion governing the evolution of an  orbit starting at $\boldsymbol{X}(0)$, along with its variational equations, which govern the evolution of a small perturbation $\boldsymbol{W}(0)=\delta \boldsymbol{X}(0)$ from this orbit~\cite{SG2010,SS2018,DMMS2019} are
\begin{equation}
    \dot{\boldsymbol{Z}}=\boldsymbol{L}_{\mathcal{HV}}\boldsymbol{Z} = \left(\boldsymbol{L}_{\mathcal{AV}} + \boldsymbol{L}_{\mathcal{BV}}\right)\boldsymbol{Z},
    \label{eq:formal_eqs_motion_var_eqs}
\end{equation}
where $\boldsymbol{Z}=\left(\boldsymbol{X}, \delta \boldsymbol{X}\right)$,  $\left(\dot{ }\right)$ denotes the time derivative and $\boldsymbol{L}_{\mathcal{HV}}$ is a Lie operator whose general expression can, for example, be found in~\cite{SG2010,DMMS2019}.
Therefore, the solution of the system's dynamical equations [Eq.~\eqref{eq:formal_eqs_motion_var_eqs}] reads 
\begin{equation}
    \boldsymbol{Z} (\tau)= e^{ \tau \left(\boldsymbol{L}_\mathcal{AV} + \boldsymbol{L}_\mathcal{BV} \right)} \boldsymbol{Z} (0).
    \label{eq:formal_solution_eqs_motion_var_eqs}
\end{equation}

A symplectic integrator consists of approximating the action of the Lie operator $e^{ \tau \left(\boldsymbol{L}_\mathcal{AV} + \boldsymbol{L}_\mathcal{BV} \right)}$ by a product of subsequent actions of operators $e^{a_i\tau \boldsymbol{L}_\mathcal{AV}}$ and $e^{b_i\tau \boldsymbol{L}_\mathcal{BV}}$ for appropriately chosen  sets of real coefficients $a_i$, $b_i$  to achieve a certain accuracy~\cite{Y1990,H2006}.
The later Lie operators can be analytically found to be
\begin{equation}
    e^{\tau \boldsymbol{L}_\mathcal{AV}} \coloneqq
    \begin{cases}
        x_j^\prime = x_j + \tau p_j & \\
        p_j^\prime = p_j  \\
        \delta x_j^\prime = \delta x_j + \tau \delta p_j & \\
        \delta p_j^\prime = \delta p_j & \\
    \end{cases},
    \quad \mbox{for $1\leq j \leq n$}
    \label{eq:lie_operator_a_part}
\end{equation}
and 
\begin{widetext}
\begin{equation}
     e^{\tau \boldsymbol{L}_\mathcal{BV}} \coloneqq
    \begin{cases}
        x _j^\prime = x_j & \mbox{for $1\leq j \leq n$} \\
        p _1^\prime = p_1 + \tau \left[ - \gamma _0 x_1 - \Gamma x_1^3 - \left(1 + \gamma\right)x_1 + \left(1 - \gamma\right)\left(x_2 - x_1\right)\right] & \\
        p _j^\prime = p _j + \tau \left[ - \gamma_0 x_j - \Gamma x_j^3 - \left(1 + \gamma\right)\left(x_j - x_{j - 1}\right) + \left(1 - \gamma\right)\left(x_{j + 1} - x_j\right)\right] & \mbox{if $j$ odd, $2\leq j \leq n - 1$} \\
        p _j^\prime = p _j + \tau \left[ - \gamma _0 x_j - \Gamma x_j^3 - \left(1 - \gamma\right)\left(x_j - x_{j - 1}\right) + \left(1 + \gamma\right)\left(x_{j + 1} - x_j\right) \right] & \mbox{if $j$ even, $2\leq j \leq n - 1$}\\
        p _n^\prime = p _n + \tau \left[ - \gamma _0 x_n - \Gamma x_n ^3 - \left(1 + \gamma\right)\left(x_n - x_{n - 1}\right)\right] & \mbox{if $n$ odd} \\
        p _n^\prime = p _n + \tau \left[ - \gamma_0 x_n - \Gamma x_n ^3 - \left(1 - \gamma\right)\left(x_n - x_{n - 1}\right)\right] & \mbox{if $n$ even} \\
        \delta x _j ^\prime = \delta x_j & \mbox{for $1\leq j \leq n$} \\
        \delta p _1 ^\prime = \delta p _1 + \tau \left[ \sigma_1 \delta x_1 + \sigma _2 \delta x_2\right] & \\
        \delta p _j ^\prime = \delta p _j + \tau \left[ \sigma _{j - 1} \delta x_{j - 1} + \sigma_j \delta x_j + \sigma_{j + 1}\delta x_{j + 1}\right]   & \mbox{if $j$ odd, $2\leq j \leq n - 1$} \\
        \delta p _j ^\prime = \delta p _j + \tau \left[ \beta_{j - 1}\delta x_{j - 1} + \beta_j \delta x_j + \beta _{j + 1} \delta x_{j + 1}\right] & \mbox{if $j$ even, $2\leq j \leq n - 1$} \\ 
        \delta p _n ^\prime = \delta p _n + \tau \left[ \sigma _{n - 1}\delta x_{n - 1} + \sigma _n \delta x_n\right] & \mbox{if $n$ odd} \\
        \delta p _n ^\prime = \delta p _n + \tau \left[ \beta_{n - 1}\delta x_{n - 1} + \beta_n \delta x_n \right] & \mbox{if $n$ even}\\
    \end{cases}
\end{equation}
\end{widetext}
where 
$\sigma _1 = -\gamma_0 - 3\Gamma x_1^2 - 2$, $\sigma _2 = 1 - \gamma$ 
%(also equal to $\sigma_{j + 1}$ with $j = 1$),
$\sigma_{j - 1} = 1 + \gamma$, $\sigma _{j + 1} = 1 - \gamma$, $\sigma_j = - \gamma _0 - 3\Gamma x_j^2 - 2$, 
$\sigma_{n - 1} = 1 +\gamma$ 
%(also obtained using $\sigma_{j - 1}$ with $j = n$), 
$\sigma _n = -\gamma _0 - 3\Gamma x_n^2 - \left(1 + \gamma\right)$,
and
$\beta _{j - 1} = 1 - \gamma$, $\beta_{j + 1} = 1 + \gamma$, $\beta _j = -\gamma_0 - 3\Gamma x_j^2 - 2$
$\beta_{n - 1} = 1 - \gamma$ 
%(equivalent to $\beta_{j - 1}$ when $j = n$), 
$\beta _n = - \gamma _0 - 3\Gamma x_n^2 - \left(1 - \gamma\right)$.

We implemented in this work the so-called $\mathcal{ABA}864$ symplectic scheme of order $4$~\cite{BCFLMM2013,FLBCMM2013} which has proved to be a very efficient integration scheme for  1D lattice Hamiltonian systems~\cite{SS2018,DMMS2019}.

%%%%%%%%%%%%%%%%%%%%%%%%%%%%%%%%%%%%%%%%
\section{\label{app-sec:floquet_resonance} The connection between the Floquet analysis and the resonant modes in the weak nonlinear limit}
\begin{figure}
    \centering
    \includegraphics[width=1\columnwidth]{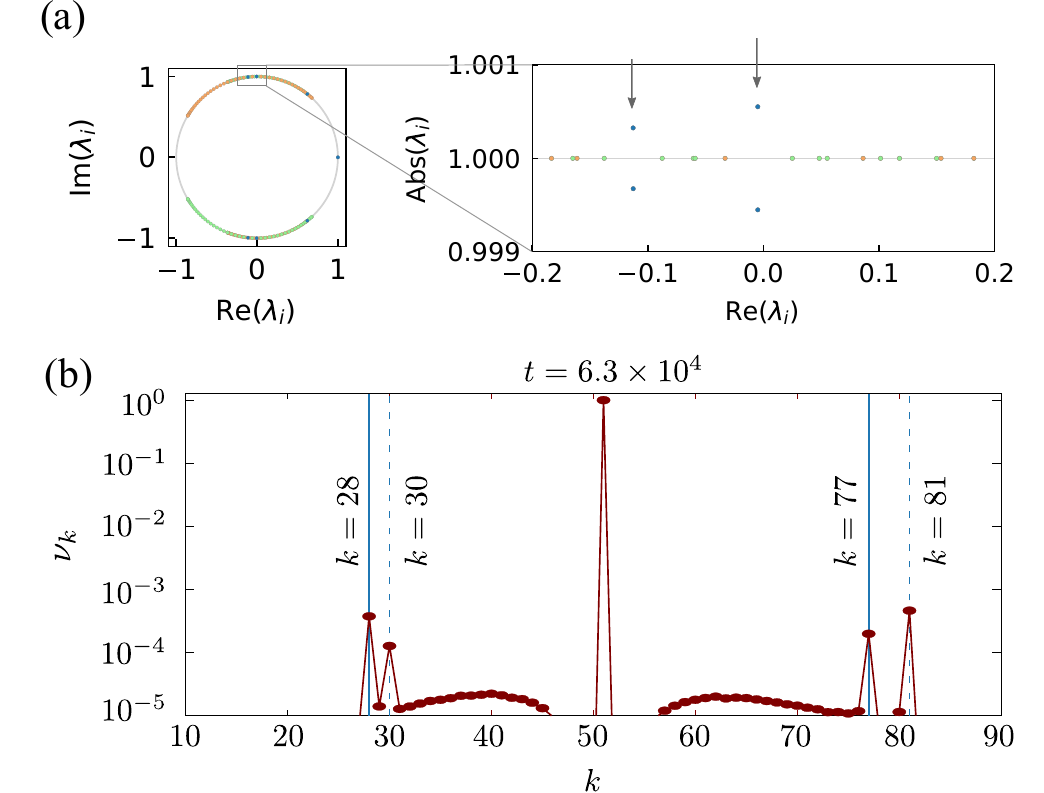}
    \caption{(a) Geometric representation of the FMs $\lambda_i$ (see text for details) in the complex plane for the topological edge state with parameter $(\mathcal{H}=0.307, \Gamma = -0.8)$ of Fig.~\ref{fig:distr_colormap_energy_mode_unstable}(a).
    We superimpose the unit circle to guide the eye.
    The upper canvas depicts a zoom into the region of the unit circle where the strongest divergence of the FMs from the unit circle  takes place. The most unstable eigenvalues are denoted by arrows.
    (b) Normalized modal energy $\nu_k(t)$ [Eq.~\eqref{eq:modal_energy}] profile at time $t\simeq 6.3\times 10^{4}$ time units for a representative realization of perturbation of the edge breather mode at $(\mathcal{H}=0.307, \Gamma=-0.8)$.
    This figure corresponds to a vertical cross section of $\nu_k(t)$ at $(t\approx 6.3\times 10^{4}, k)$, see white vertical line of Fig.~\ref{fig:distr_colormap_energy_mode_unstable}(a). 
    The vertical bold and dotted lines label the resonant modes.
    }
    \label{fig:floquet_multipliers}
\end{figure}

We identify the frequencies of the most unstable eigenvectors $v_i$ associated with the Floquet multipliers (FMs) $\lambda_i$ which diverge the farthest from the unit circle and map these frequencies to the first resonant modes responsible of the lattice thermalization in the weak nonlinear regime.
In practice, the linear stability of a periodic orbit $\boldsymbol{X}(0)$ with period $T$ is estimated by following the time evolution of a small perturbation $\boldsymbol{W}(0)$ to $\boldsymbol{X}(0)$ (see also Sec.~\ref{subsec:stochastic_threshold}).
The temporal evolution of such perturbation can be expressed as 
\begin{equation}
    \boldsymbol{W}(t)= \boldsymbol{A}(t)\cdot W(0),
\end{equation}
where $\boldsymbol{W}(t)$ is the pertubation at time $t>0$ and $\boldsymbol{A}(t)$ is the {\it fundamental matrix} of the system's variational equations (see e.g.~\cite[App.~B]{CXYKT2021} and~\cite{S2001} for further details).
It follows that the values of the perturbation after a time period $t=T$ is
\begin{equation}
    \boldsymbol{W}(T) = \boldsymbol{M}\cdot \boldsymbol{W}(0), \quad \boldsymbol{M} = \boldsymbol{A}(T),
\end{equation}
in which $\boldsymbol{M}$ is called {\it monodromy matrix}.
Consequently, the stability properties of the periodic orbit $\boldsymbol{X}(0)$ are encompassed within the eigen-characteristics of $\boldsymbol{M}$ (see e.g.~\cite{S2001,K2019,CCQ2020}). 
The $2n$ eigenvalues of $\boldsymbol{M}$ are referred to as FMs, $\lambda_i$ and are associated with $2n$ eigenvector $\boldsymbol{v}_i$.
Whether any of the $\lvert \lambda_i\rvert \neq 1$ (diverges from the unit circle in the complex plane), the periodic orbit $\boldsymbol{X}(0)$ is said to be unstable.

In Fig.~\ref{fig:floquet_multipliers}(a), we show the $\lambda_i$ for the edge breather mode with parameter $(\mathcal{H}=0.307, \Gamma=-0.8)$ and observe the presence $4$ FMs (blue dots highlighted inside the box) that lead to the first two most dominant instabilities. These instabilities (in blue) result from the collision of two eigenvalues with opposite Krein signature denoted by orange and green colors.
Next, we can co-relate these colliding eigenvalues to the eigenvalues of the linear dispersion band (details can be found in~\cite[App.~B]{CXYKT2021}). 
We find that the appearance of instability originates from the collisions between the modes $k=28$ [resp. $k=30$] and $k=81$ [resp. $k=77$] of the acoustic and optical bands.

Figure~\ref{fig:floquet_multipliers}(b) show the normalized energy per mode $\nu_k$ [Eq.~\eqref{eq:modal_energy}] at time $t\simeq 6.3\times 10^{4}$ for a representative simulation using as initial condition the perturbation of the topological edge state with $(\mathcal{H}=0.307, \Gamma=-0.8)$ of Figs.~\ref{fig:distr_colormap_energy_mode_unstable}(a) and~\ref{fig:floquet_multipliers}.
We clearly see that the first $4$ resonant modes correspond to the frequencies of the most unstable eigenvectors $\boldsymbol{v}_i$ of the Floquet analysis.

\section{\label{app-sec:width_resonance} Width $\chi _k$ and frequency shift as a function of energy}

\begin{figure}[!]
     \centering
     \includegraphics[width=1\columnwidth, height=0.8\linewidth]{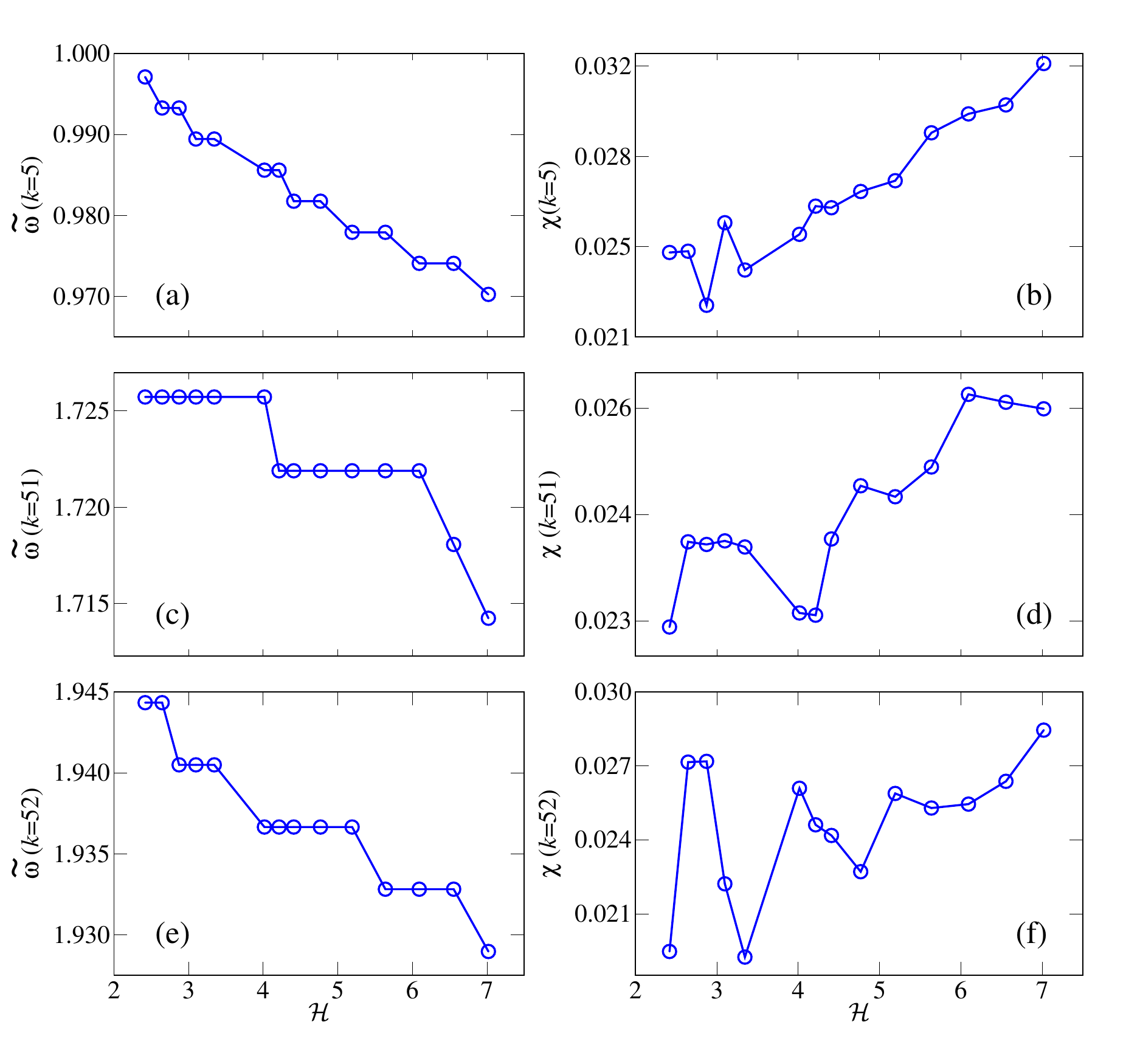}
     \caption{Frequency shift, $\tilde{\omega}_k=\tilde{\omega}(k)$ [Eq.~\eqref{eq:renorm_disper_relation}] [panels (a), (c), (e)] and width of the frequency shift distribution, $\chi_k=\chi (k)$ [Eq.~\eqref{eq:width_freq_distr}] [panels (b), (d), (f)]  as functions of the total energy $\mathcal{H}$ [Eq.~\eqref{eq:hamilton_tkg}] of the system: (a-b) $k=5$, (c-d) $k=51$ and (e-f) $k=52$.
     The ground spring nonlinear coefficient $\Gamma$, is fixed at $\Gamma=-0.8$. 
    %  \textcolor{green}{[HS: The vertical axis in the second column is $\chi$ or $\chi_k$?]}
     }
     \label{fig:frequency_shift_&_width_vs_energy}
\end{figure}

\begin{figure}[!]
    \centering
    \includegraphics[width=0.75\columnwidth]{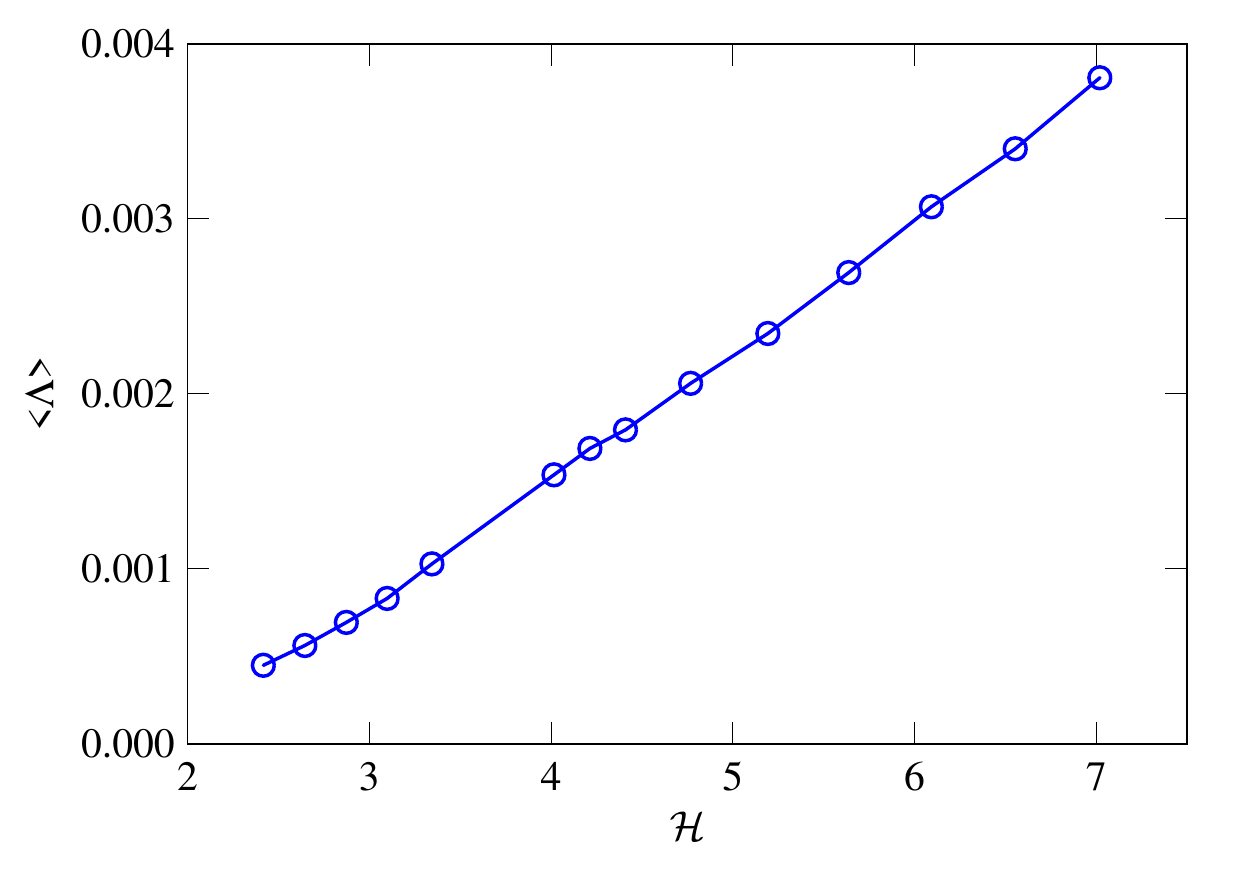}
    \caption{The MLE, $\langle \Lambda\rangle$ as function of the energy of the system $\mathcal{H}$ [Eq.~\eqref{eq:hamilton_tkg}].
    The value of $\langle \Lambda\rangle$ grows almost linearly with $\mathcal{H}$.
    We numerically approximated $\langle \Lambda \rangle$ by $\langle \lambda (T)\rangle$,
    % [HS: This has to be clearly stated also in the etx. As mentioned earlier we have to be careful on our notations related to LEs, what we compute, what we present and what we estimate.]} 
    the ftMLE at the end of the integration time $T$ for each set of parameters in the lattice.
    The parameters used in $\mathcal{H}$ [Eq.~\eqref{eq:hamilton_tkg}] are similar to the ones of Fig.~\ref{fig:frequency_shift_&_width_vs_energy}.
    }
    \label{fig:mle_ft_energy}
\end{figure}

To examine the role of the system's nonlinearity
% \textcolor{green}{[HS: no what?]}, 
% like the energy dependence in the frequency shift, 
we quantify the width $\chi _k$ of the frequency shift distribution~\cite{LO2018} 
 \begin{equation}
    \chi _k = \sqrt{\sum_{\Omega}\left(\Omega - \tilde{\omega}_k \right)^2 A_{\Omega, k}}, \quad A_{\Omega, k} = \frac{\langle \lvert a_k (\Omega)\rvert^2\rangle}{\sum _{\Omega} \langle \lvert a_k (\Omega)\rvert^2\rangle},
    \label{eq:width_freq_distr}
 \end{equation}
where $\Omega$ and $\tilde{\omega}_k$ [Eq.~\eqref{eq:renorm_disper_relation}] are the frequency and renormalized frequency of the $k$th mode.
We calculate the frequency shift $\tilde{\omega}_k$ [Eq.~\eqref{eq:renorm_disper_relation}] and the broadening of the frequency shift distribution, $\chi_k$ [Eq.~\eqref{eq:width_freq_distr}]
% \textcolor{green}{[HS: How exacly is this evaluated?]} 
for different values of the energy $\mathcal{H}$ [Eq.~\eqref{eq:hamilton_tkg}], having as an initial condition always the topological linearly unstable nonlinear mode, as shown in Fig.~\ref{fig:frequency_shift_&_width_vs_energy}.
The results are presented for a mode  of the lower part of the spectrum ($k=5$) in  Figs.~\ref{fig:frequency_shift_&_width_vs_energy}(a), (b), for the mode $k=51$ at the center of the frequency band in Figs.~\ref{fig:frequency_shift_&_width_vs_energy}(c), (d) and for a mode at the upper part of the spectrum of frequencies ($k=52$) in Figs.~\ref{fig:frequency_shift_&_width_vs_energy}(e), (f).
Upon increasing the value of $\mathcal{H}$, the $\tilde{\omega}_k$ is decreasing and the $\chi_k$ is, in general, growing for all  mode numbers. The latter observable, is positively correlated with the system's MLE, $\langle\Lambda\rangle$ [Eq.~\eqref{eq:mle}] which is linearly increasing, Fig.~\ref{fig:mle_ft_energy}.
The somehow non-smooth nature of the trends of $\tilde{\omega}_k$ and $\chi_k$ (Fig.~\ref{fig:frequency_shift_&_width_vs_energy}) can be attributed to the fact that we are exploring small values of the effective nonlinear parameter $\sim \lvert \Gamma \mathcal{H}_2(t=0)/n \rvert \lessapprox 0.006$~\cite{OVPL2015,LO2018,PCBLO2019} for which mild nonlinear effects are experienced by the system.
For instance a small degree of chaos, $\langle \Lambda \rangle \lessapprox 0.004$, seeing in Fig.~\ref{fig:mle_ft_energy}.
% \textbf{GT: Figure 9 is not described in details. We can discuss, but I think figures 8 and 9 could move into the Appendix.}
Nevertheless, it can be conjectured that as the value of the energy increases, the frequency overlap between modes is becoming wider, leading to more prominent resonances between NMs. 
 This results to a faster decay of the energy of the initially excited modes $a_k^{b} (t=0)$, including the one that corresponds to the edge mode ($k=51$) in which most of the initial energy is located.
%   \textbf{GT: including the one that corresponds to the edge mode ($k=51$) in which most of the initial energy is located.}
  %which is both spatially localized and protected by the large frequency band %gap. 


%%%%%%%%%%%%%%%%%%%%%%%%%%
%%%%%%%%%%%%%%%%%%%%%%%%%%
%\nocite{*}
% \bibliographystyle{apsrev4}
\let\itshape\upshape
\normalem
\bibliography{references}% Produces the bibliography via BibTeX.

\end{document}